\documentclass[11pt,a4paper]{article}

\usepackage{amssymb}
\usepackage{amsmath}
\usepackage{amsfonts}
\usepackage{epsfig}
\usepackage{graphicx}
\usepackage{color}
\usepackage{bigints}
\usepackage{array}
\usepackage{dsfont}
\usepackage{subcaption}
\usepackage{amsthm}
\usepackage{hyperref}

\usepackage{authblk}

\usepackage{algorithm}
\usepackage{algorithmic}

\newtheorem{definition}{Definition}[section]
\newtheorem{theorem}{Theorem}[section]
\newtheorem{lemma}{Lemma}[section]
\newtheorem{corollary}{Corollary}[section]

\allowdisplaybreaks

\usepackage{ifthen}
\newboolean{long}
\setboolean{long}{true}

\begin{document}

\title{Frequency Estimation Under Multiparty Differential Privacy: \\ One-shot and Streaming}


\author[1]{Ziyue Huang}
\author[1]{Yuan Qiu}
\author[1]{Ke Yi}
\author[2]{Graham Cormode}

\affil[1]{Hong Kong University of Science and Technology, \texttt{\{zhuangbq,yqiuac,yike\}@cse.ust.hk}}
\affil[2]{University of Warwick, \texttt{g.cormode@warwick.ac.uk}}

\date{}

\maketitle

\begin{abstract}
We study the fundamental problem of frequency estimation under both privacy and communication constraints, where the data is distributed among $k$ parties. We consider two application scenarios: (1) one-shot, where the data is static and the aggregator conducts a one-time computation; and (2) streaming, where each party receives a stream of items over time and the aggregator continuously monitors the frequencies. We adopt the model of multiparty differential privacy (MDP), which is more general than local differential privacy (LDP) and (centralized) differential privacy. Our protocols achieve optimality (up to logarithmic factors) permissible by the more stringent of the two constraints. In particular, when specialized to the $\varepsilon$-LDP model, our protocol achieves an error of $\sqrt{k}/(e^{\Theta(\varepsilon)}-1)$ using $O(k\max\{ \varepsilon, \frac{1}{\varepsilon} \})$ bits of communication and $O(k \log u)$ bits of public randomness, where $u$ is the size of the domain.
\end{abstract}

\section{Introduction}
Consider a distributed system with $k$ parties, where each party has some data. We study two settings for privacy-preserving data analysis where an aggregator wants to obtain some aggregated statistics over all data held by the parties: (1) one-shot, where the data is static and the parties conduct a one-time computation; and (2) streaming, where each party receives a stream of items over time and the aggregator wants to monitor the aggregated statistics continuously. These settings are motivated by real-world applications. For example, where each hospital holds some medical records which some third-party public health organization wishes to analyze, and it is the responsibility of the hospital to protect the privacy of their patients. In addition to the privacy constraint, the communication cost often becomes a major bottleneck which limits the scalability of distributed machine learning tasks. For instance, in federated learning, the data is distributed among many mobile devices connected by bandwidth-limited wireless links and the communication is expensive and unreliable \cite{kairouz2019advances}. This poses a dual-challenge to the design of the protocol.   Recently, there has been a lot of interest in designing protocols under both constraints \cite{wang2017locally, chen2020breaking, ghazi2020private, agarwal2018cpsgd, acharya2019hadamard, acharya2019communication}. The most important result obtained in this paper is that, for the frequency estimation problem, it is possible to achieve optimality permissible by the more stringent of the two constraints.

\subsection{Problem Formulation}
\subsubsection{Multiparty Differential Privacy}
In the settings above, the standard notion of privacy is \textit{multiparty differential privacy (MDP)} \cite{beimel2008distributed, pathak2010multiparty, hamm2016learning}.   Denote the parties as $P_1,\dots, P_k$, and the aggregator as $P_0$. Suppose each party $P_i$, $i\ge 1$, has a multiset $\mathcal{D}_i$ of $n_i \ge 1$ items, drawn from a universe $\mathcal{U}$ of size $u$.  Let $N = \sum_{i=1}^k n_i$. We use $[n]$ to denote $\{1,\dots,n\}$.

\begin{definition}[Multiparty Differential Privacy \cite{vadhan2017complexity}]
Let $P$ be a protocol involving parties $(P_0, P_1, \ldots, P_k)$, where $P_i$ has input dataset $\mathcal{D}_i \in \mathcal{U}^{n_i}, i\in [k]$, while $P_0$ has no input.  Consider any party $P_i$, $i=0,1,\dots,k$, and let $A$ be an adversary controlling $P_{-i} = \{P_0,\dots, P_k\}-\{P_i\}$. We use $\operatorname{View}_{P_{-i}} ( P_{-i} \leftrightarrow (P_0, \ldots, P_k) (\mathcal{D}) )$ to denote the random variable that includes everything that $A$ sees when participating in the protocol on input dataset $\mathcal{D}=(\mathcal{D}_1,\dots, \mathcal{D}_k)$.  We say that $P$ is $\varepsilon$-differentially private if for every $i\in [k]$ and every two neighboring datasets $\mathcal{D}, \mathcal{D}' \in (\mathcal{U}^{n_1}, \ldots, \mathcal{U}^{n_k})$ that differ on one item in $P_i$'s input, the following holds for every set $T$:
\begin{align*}
& \Pr [ \operatorname{View}_{P_{-i}} ( P_{-i} \leftrightarrow (P_0, \ldots, P_k) (\mathcal{D}) ) \in T ]  \le e^\varepsilon \cdot \Pr [ \operatorname{View}_{P_{-i}} ( P_{-i} \leftrightarrow (P_0, \ldots, P_k) (\mathcal{D}') ) \in T ].
\end{align*}
\end{definition}
Note that the MDP model is more general than other popular privacy models such as \textit{local differential privacy (LDP)} and \textit{(centralized) differential privacy (DP)}: Setting $n_i=1$ for all $i \in [k]$ yields the former while setting $k=1$ yields the latter.

\ifthenelse{\boolean{long}}{
Depending on the power of the adversary and the structure of the communication graph, we arrive at different variants of the model.  Clearly, more powerful adversaries imply stronger algorithmic results but weaker lower bounds.  In this paper, all protocols guarantee privacy even with active adversaries (i.e., they may alter the behavior of the parties to deviate from the specified protocol), while the lower bounds hold for passive adversaries (also known as ``semi-honest'' parties, i.e., the adversary observes the transcript but cannot change it). Note that, however, we can no longer guarantee the utility of the protocol when the adversary is active, as such an adversary can completely change the parties' inputs.  We also comment that the MDP definition above is information-theoretic, i.e., it assumes that the adversary has infinite computing power.  It is also possible to consider computationally bounded adversaries \cite{beimel2008distributed, dwork2006our}, but we will not explore that direction in this paper.  

Similarly, more restrictive communication graphs lead to more general protocols.  All protocols in this paper work in the most restrictive, \textit{one-way client-server} model, i.e., all messages are sent by the clients (parties) to the server (aggregator). In fact, it is this restrictive communication pattern that makes the protocols insensitive to the type of the adversary (active or passive). On the other hand, the lower bounds we match hold even in the \textit{blackboard} model, in which every message sent to the aggregator is visible to all parties at no additional cost.
}{}


\subsubsection{Frequency Estimation}
Denote the frequency of item $j\in [u]$ on party $i$ by $x_{i,j}$, which we call the \textit{local count}.  In the \textit{frequency estimation} problem, the aggregator wishes to obtain a synopsis, which can be used to extract an estimate of the \textit{global count} $y_j = \sum_{i=1}^k x_{i,j}$ for any $j\in [u]$.  As with prior work \cite{cormode2005improved, wang2017locally, bassily2017practical}, we aim at an additive error guarantee that holds for a single query with probability $1-\beta$.  A vectorized view of the problem is to consider the local counts $\{x_{i,j}\}_j$ at party $i$  as a vector $\mathbf{x}_i \in \mathbb{N}^u$, and we want to obtain a $\tilde{\mathbf{y}}$ that minimizes $\|\tilde{\mathbf{y}} - \mathbf{y}\|_\infty$, where $\mathbf{y}=\sum_i \mathbf{x}_i$. Setting $\beta=O(1/u)$ plus a union bound converts any error guarantee of the former into one of the latter.

\begin{figure*}[htbp]
     \centering
     \includegraphics[scale=0.9]{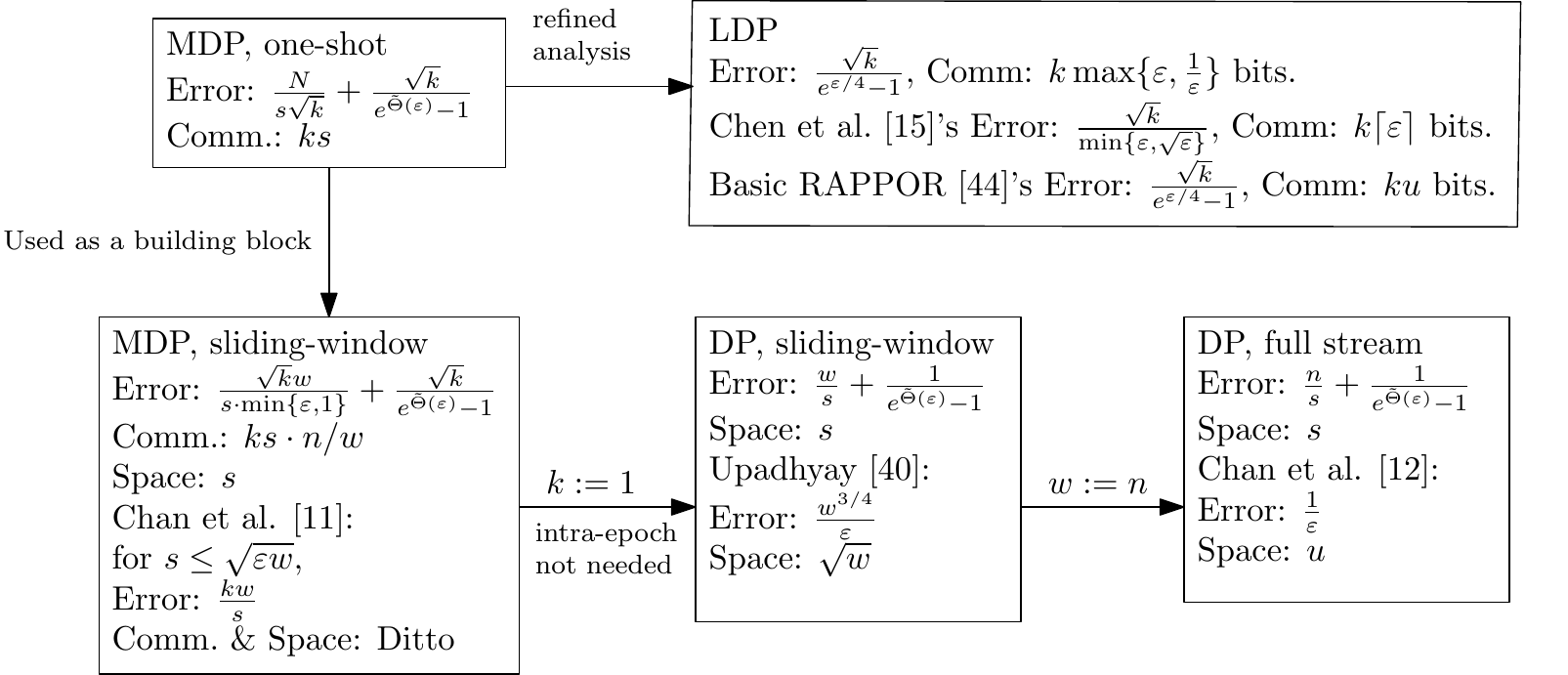}
     \caption{Overview of our results, where $s$ can be any positive integer.  All bounds suppress $\log^{O(1)}(N/\beta)$ factors, and the communication is measured in $O(\log (uN/(\beta\varepsilon)))$-bit words.}
     \label{fig:result}
\end{figure*}

Figure~\ref{fig:result} gives an overview of our results, wherein we omit $\log^{O(1)}(N/\beta)$ factors.  In the next two subsections, we elaborate on these results and compare them with prior work.  Optimality also holds subject to these polylogarithmic factors.  Communication will be measured in $O(\log (uN/(\beta\varepsilon)))$-bit words; when measured in bits, we will state so explicitly, in which case logarithmic factors must not be omitted.  

Besides applications of its own, a frequency estimation protocol can be used as a building block to solve many related problems such as heavy hitters, quantiles, and orthogonal range counting (in constant dimensions), at the cost of some extra polylogarithmic factors \cite{cormode2010methods, bassily2017practical, cormode2019answering, wang2013quantiles}, using reductions that are now considered standard.  

\subsection{Our Results and Prior Work}

\subsubsection{One-shot Protocols}

Given any $\varepsilon > 0$ and $s \ge 1$, our one-shot frequency estimation protocol achieves an error of $\tilde{O}(N/(\sqrt{k}s)) + \sqrt{k}/(e^{\tilde{\Theta}(\varepsilon)}-1)$ with $\tilde{O}(ks)$ communication.  This is optimal for the high privacy regime $\varepsilon=O(1)$.  To see this, observe that the first error term is communication-dependent while the second term is privacy-bound. The communication-bound term matches the lower bound in the non-private, blackboard communication model for all $1\le s \le cN/k$ where $c$ is some small constant \cite{huang2017communication}.  For $s>cN/k$, the communication-bound error term becomes $\tilde{O}(\sqrt{k})$, which is dominated by the privacy-bound term for $\varepsilon=O(1)$.  Meanwhile, it is known that regardless of the communication cost, the error has to be $\tilde{\Omega}(\sqrt{k}/\varepsilon)$ for the high privacy regime $\varepsilon =O(1)$ due to the privacy constraint \cite{bassily2015local}, even if each party has just one item (i.e., the LDP model). 



In the LDP model, the RAPPOR protocol (the basic version) \cite{wang2017locally, erlingsson2014rappor} achieves an error $O(\sqrt{k}/(e^{\varepsilon/4}-1))$ with a large communication cost $O(ku)$. The now classical \textit{Hadamard Randomized Response (HRR)} algorithm \cite{wang2017locally, bassily2015local} reduces the communication cost to $O(k)$ bits and uses $O(k \log u)$ bits of public randomness\footnote{These random bits can be communicated if the public randomness is not available. }.  And HRR achieves an error of $O\left(\sqrt{k \log \frac{1}{\beta}}/\min \{ \varepsilon, 1 \}\right)$, which is optimal for the high privacy regime $\varepsilon < 1$ \cite{bassily2015local}. The general privacy regime $\varepsilon = \Omega(1)$ is also of both theoretical and practical interest, which has drawn some attention recently \cite{chen2020breaking, acharya2019hadamard,kairouz2014extremal,wang2017locally}. For the general privacy regime $\varepsilon > 1$, Chen et al.~\cite{chen2020breaking}\footnote{They also study the $\ell_1/\ell_2$ error guarantees which are not considered in this paper.} show that it is possible to reduce the error to $O\left(\sqrt{k \log\frac{1}{\beta} / \varepsilon }\right)$, while using $O(k \varepsilon)$ bits of communication and $O(k\varepsilon \log u)$ bits of public randomness. A more refined analysis shows that when specialized to the LDP model (i.e., $n_i=1$ for all $i$, hence $k=N$),
our protocol 
achieves an error of  $O\left(\sqrt{k\log\frac{1}{\beta}}/(e^{\varepsilon/4}-1) + \log\frac{1}{\beta} \right)$ with $O(k \varepsilon)$ bits of communication and $O(k\log u)$ bits of public randomness. This makes an exponential improvement in $\varepsilon$ over \cite{chen2020breaking} for $\varepsilon > 1$ while using less public randomness.  Finally, we present a lower bound of $\Omega(\sqrt{k} / e^{\varepsilon/2})$ for $\varepsilon = O(\log k)$, regardless of communication cost.  This improves the previous lower bound $\tilde{\Omega}(k^{1/3})$ for $\varepsilon = O(\log k)$ in \cite{ghazi2019power}. In addition, combined with the existing lower bound $\Omega(\sqrt{k}/\varepsilon)$ for $\varepsilon = O(1)$ in \cite{bassily2015local}, this implies that the error of our LDP protocol is optimal for $\varepsilon = O(\log k)$, up to a constant-factor difference in $\varepsilon$.

\subsubsection{Streaming Protocols}

The MDP model has a natural streaming version. As with prior work \cite{chan2012differentially,cormode12continuous,cormode11functional}, we adopt a \textit{synchronous} timing model, where time is divided into discrete steps, and one item arrives at each party in every time step.  Messages sent within a time step all arrive before the next time step.  In practice, the parties' clocks might be out-of-sync and messages can be delayed.  In this case, we can include timestamps in the messages to simulate the execution, which is particularly easy for our protocol as it only uses one-way messages from the parties to the aggregator.  The assumption that one item must arrive at each time step is also without loss of generality.  If nothing arrives at a party in a time step, this can be treated as a dummy item. 

Perhaps the most useful streaming model is the \textit{sliding-window} model.  Here, the goal is for the aggregator to maintain a synopsis such that the global count of any item $j$, counting all items that have arrived in the last $w$ time steps, can be estimated. However, all messages sent during the entire streaming period, not just those sent in the sliding window, must collectively be $\varepsilon$-DP as in the one-shot MDP model.  We present an $\varepsilon$-MDP protocol for this problem that, for any integer $s\ge 1$, achieves error $\tilde{O}(w\sqrt{k}/(s \cdot \min \{ \varepsilon, 1 \})) + \sqrt{k}/(e^{\tilde{\Theta}(\varepsilon)}-1)$ with $\tilde{O}(ks \cdot n/w)$ communication, while using $\tilde{O}(s)$ space on each party, where $n$ is the total number of time steps.  Note that the one-shot problem is a special case of this problem, by just setting $n:=w, N:=kw$, and asking for the synopsis only at the end of the stream.  Compared with our one-shot result, the only difference is that the communication-bound error term has an extra $1/\min\{\varepsilon,1\}$ factor.  Thus, the communication-error trade-off of our sliding window protocol is still optimal for $\varepsilon=\Omega(1)$, but optimality is open for $\varepsilon=o(1)$.   For $\varepsilon=\Theta(1)$, the previous result for this problem \cite{chan2012differentially} gave an error of $\tilde{O}(kw/s)$ under the same communication budget, which is a $\sqrt{k}$-factor from optimal. More importantly, the largest $s$ that can be supported by the protocol of \cite{chan2012differentially} is $O(\sqrt{w})$, which means that the minimum error achievable (regardless of communication) is $\tilde{O}(k\sqrt{w})$. This is a $\sqrt{kw}$-factor from $\tilde{O}(\sqrt{k})$, the smallest error permissible by the privacy constraint, which can be achieved by our protocol by setting $s:=w$. 

For (private or non-private) streaming algorithms, an important measure of complexity is space.  To see that our space-error trade-off is also optimal, simply consider the degenerate case where $k=1$.  This particular case has actually been recently studied by \cite{upadhyay2019sublinear}, where an algorithm with error $\tilde{O}(w^{3/4}/\varepsilon)$ and space $\tilde{O}(\sqrt{w})$ is presented. When degenerated to the $k=1$ case, our protocol achieves $\tilde{O}({w\over s}) + 1/(e^{\tilde{\Theta}(\varepsilon)}-1)$ error with $\tilde{O}(s)$ space.  To compare with \cite{upadhyay2019sublinear}, just set $s=\sqrt{w}$, which yields $\tilde{O}(\sqrt{w}) + 1/(e^{\tilde{\Theta}(\varepsilon)}-1)$ error.  In fact, the protocol in \cite{chan2012differentially} yields the same space-error trade-off as ours (but with a limited range $s\le \sqrt{\varepsilon w}$) for the $k=1$ case, which was overlooked in \cite{upadhyay2019sublinear}.  
In an analogy to the communication-error trade-off, the space-error trade-off is also determined by space or privacy, whichever is more stringent.  The space-bound term $\tilde{O}({w \over s})$ is optimal (assuming $u\ge w$) by well-known lower bounds in the (non-private) streaming literature \cite{chakrabarti2015data}, while the optimality of the privacy-bound term follows from the centralized DP lower bound $\tilde{\Omega}(\frac{1}{\varepsilon})$ for $\varepsilon = O(\log u)$, even for the one-shot problem \cite{vadhan2017complexity}.
Thus, our protocol achieves the optimal space-error trade-off for the full range $1\le s \le w$ for $\varepsilon = O(\log u)$.  On the other hand,  \cite{chan2012differentially} achieved optimality only for a partial range $1\le s \le \sqrt{\varepsilon w}$, while the results of \cite{upadhyay2019sublinear} are not optimal.
Finally, our protocol spends $\tilde{O}(1)$ time to process each item, so it is time-optimal as well.  

\ifthenelse{\boolean{long}}{\subsection{Other Related Work}
We briefly mention results in other models of privacy that are relevant to our study. 
%
Most closely related are the notions of \textit{continual observation}, and \textit{pan privacy}, which consider privacy against an adversary who may observe a snapshot of the algorithm's internal state (pan privacy), or when the algorithm continually publishes updates based on new readings (continual observation).  The latter of these most closely matches our (distributed) streaming results, and we leverage similar techniques, such as expressing partial aggregations within a tree structure~\cite{chan2011private,dwork2010differential}. 
More specifically, by setting $w:=n$, the sliding-window model degenerates into the problem of monitoring the full stream, i.e., continual observation.  Existing solutions for this problem achieves $\tilde{O}(1/\varepsilon)$ error with $\tilde{O}(u)$ space \cite{chan2011private}.  Our protocol achieves the same error with space $\tilde{O}(n)$, which can be much smaller than $\tilde{O}(u)$.  In addition, we provide a full-range space-error trade-off.

We have mentioned the model of Local Differential Privacy (LDP), which corresponds to MDP with $n_i=1$ for all $i$ in the one-shot setting. 
The most impactful work in this model is concerned with frequency estimation, and finding heavy hitters, based on building ``frequency oracles'' from each site's message to estimate item frequencies, as in our setting~\cite{erlingsson2014rappor,appledp,ding2017collecting,bassily2020practical}.
Further work has studied a wide variety of data analysis and ML tasks, such as multidimensional statistics, language models and classifiers; we refer the reader to surveys on this topic~\cite{wang2020survey,yang2020survey,cormode2018tutorial}. 

Most recently, additional models have been proposed which aim to achieve improved privacy-accuracy tradeoffs by making stronger assumptions. 
The shuffle model assumes that the identity of the sender can be fully disassociated from the messages sent, either by a trusted ``shuffler'' entity, or through a cryptographic mix network~\cite{erlingsson2019shuffling,balle2019privacy}. 
}{We also briefly mention results in other models of privacy that are relevant to our study in the supplementary material. 
%
Most closely related are the notions of \textit{continual observation}, and \textit{pan privacy}, which consider privacy against an adversary who may observe a snapshot of the algorithm's internal state (pan privacy), or when the algorithm continually publishes updates based on new readings (continual observation).  The latter of these most closely matches our (distributed) streaming results, and we leverage similar techniques, such as expressing partial aggregations within a tree structure~\cite{chan2011private,dwork2010differential}. 
More specifically, by setting $w:=n$, the sliding-window model degenerates into the problem of monitoring the full stream, i.e., continual observation.  Existing solutions for this problem achieves $\tilde{O}(1/\varepsilon)$ error with $\tilde{O}(u)$ space \cite{chan2011private}.  Our protocol achieves the same error with space $\tilde{O}(n)$, which can be much smaller than $\tilde{O}(u)$.  In addition, we provide a full-range space-error trade-off.
}


\section{Preliminaries}

\subsection{Differential Privacy}
The standard (centralized) differential privacy model is a special case of MDP, but we state its definition again for clarity.  Let $\mathcal{D} \sim \mathcal{D}'$ denote two neighboring datasets, which differ by one item. 
\begin{definition}[Differential Privacy \cite{dwork2014algorithmic}]
For $\varepsilon > 0$, an algorithm $\mathcal{M}$ is $\varepsilon$-differentially private (DP) if for any neighboring datasets $\mathcal{D} \sim \mathcal{D}'$ and any $S \subseteq \mathsf{Range}(\mathcal{M})$,
\begin{align*}
\mathsf{Pr}[\mathcal{M}(\mathcal{D}) \in S] \le e^\varepsilon \cdot \mathsf{Pr}[\mathcal{M}(\mathcal{D}') \in S].
\end{align*}
\end{definition}

Note that the MDP model degenerates to this definition by setting $k:=1$: $P_1$ runs $\mathcal{M}$ on $\mathcal{D}_1$ and sends $\mathcal{M}(\mathcal{D}_1)$ to $P_0$. 

For a numeric query $q$, one common DP mechanism is to add noise drawn from a symmetric geometric distribution \cite{ghosh2012universally, chan2012optimal} calibrated to $\mathrm{GS}_q := \max_{\mathcal{D} \sim \mathcal{D}'} |q(\mathcal{D}) - q(\mathcal{D}')|$, which is known as the (global) \textit{sensitivity} of $q$.

\begin{definition} [Symmetric Geometric Distribution \cite{ghosh2012universally, chan2012optimal}]
Let $\alpha > 1$. We denote by $\text{Geom}(\alpha)$ the symmetric geometric distribution that takes integer values such that the probability mass function at $l$ is $\frac{\alpha-1}{\alpha+1} \cdot \alpha^{-|l|}$.
\end{definition}

The following properties of the symmetric geometric distribution are useful.  Let $X\sim \text{Geom}(\alpha)$:
\begin{enumerate}
\item $\mathsf{E}[X]=0$; $\mathsf{Var}[X] = 2\alpha/(\alpha-1)^2 = O(1/\log^2 \alpha)$.
\item For every $d > 0$, $\Pr[|X| > d] \le \alpha^{-d}$.
\item The mechanism $\mathcal{M}(\mathcal{D}):=q(\mathcal{D}) + X$ is $\varepsilon$-DP by setting $\alpha = \exp(\varepsilon/\mathrm{GS}_q)$. Note that in this case, $\mathsf{Var}[X] = O((\mathrm{GS}_q/\varepsilon)^2)$.
\end{enumerate}

The LDP model is another degeneration of MDP by setting $n_i=1$ for all $i\in [k]$.  The \textit{Hadamard Randomized Response (HRR)} algorithm \cite{bassily2015local, wang2017locally} can be used to solve the frequency estimation problem under LDP. \ifthenelse{\boolean{long}}{Assume, without loss of generality, that $u$ is a power of 2, and recall that the Hadamard matrix can be defined recursively as
\[
H_{u} = \left[ \begin{array}{cc}
H_{u/2} & H_{u/2} \\ H_{u/2} & -H_{u/2} \end{array} \right],
\]where $H_1 = [1]$.}{} Each party $i$ samples an index $r_i$ u.a.r. from $[u]$,  encodes her item $v_i$ into a single bit $H[r_i, v_i]$, and then sends it to the aggregator via randomized response. Specifically, each user sends a one-bit message $\mathcal{M}(v_i)$ to the aggregator (if there is no shared randomness, the random index $r_i$ should also be sent using $\log u$ bits), where
\[
\mathcal{M}(v_i) = \begin{cases} 
H[r_i, v_i], & \text{w.p.  } \frac{e^{\varepsilon}}{e^{\varepsilon}+1}; \\
-H[r_i, v_i], & \text{otherwise. }
\end{cases}
\]The frequency estimator (at the aggregator side) for any item $v$ is $\frac{e^\varepsilon
 + 1}{e^\varepsilon
 - 1} \sum_{i=1}^k \mathcal{M}(v_i) \cdot H[r_i, v]$.
The error guarantee of HRR is $O(\sqrt{k \log (1/\beta)}/\min \{ \varepsilon, 1 \})$ which holds for a single query with probability $1-\beta$.

Note that the MDP/LDP model allows arbitrary interactions among the parties, and the lower bounds~\cite{bassily2015local} hold under this setting.  However, most of existing protocols (including ours) use one-way messages, except for broadcasting some public parameters to all parties before the protocol starts.  In this case, it is sufficient for each party to run a $\varepsilon$-DP mechanism $\mathcal{M}$ on her dataset $\mathcal{D}_i$ and send $\mathcal{M}(\mathcal{D}_i)$ to the aggregator.
The resulting protocol then trivially satisfies MDP/LDP against active adversaries.  It is worth pointing out that one can relax the MDP model by only allowing the adversary to control a smaller number of parties.  In this case, one may achieve errors lower than the LDP lower bound of $\Omega(\sqrt{k}/\varepsilon)$ with interactive protocols \cite{shi17coalition}.

\subsection{Count Sketch}
The \textit{count sketch} \cite{charikar2002finding} of a vector $\mathbf{x}$ of size $u$ is another vector $C(\mathbf{x})$ of size $s$,
\begin{align*}
C(\mathbf{x})[j] = \sum_{i \in [u]: h(i) = j} g(i) x_i, \quad j = 1, \ldots, s,
\end{align*}
where $h : [u] \rightarrow [s]$ and $g : [u] \rightarrow \{ -1, +1 \}$ are two hash functions.  For our analysis, we assume $h$ is pairwise-independent while $g$ is truly random.  In some cases, the latter assumption can also be relaxed to pairwise-independence.

The count sketch can be used to extract \textit{point estimates}. For any $i\in [u]$, an estimator for $x_i$ is $\tilde{x}_i = g(i) \cdot C(\mathbf{x})[h(i)]$.  It is known that $\mathsf{E}[\tilde{x}_i] = x_i$ and $\mathsf{Var} [ \tilde{x}_i ] \le \| \mathbf{x} \|_2^2 / s$.  So by the Chebyshev inequality, the error $|\tilde{x}_i - x_i|$ is $O(\| \mathbf{x} \|_2 / \sqrt{s})$ with constant probability, which is an $\ell_2$ error guarantee. Meanwhile, the count-sketch also enjoys an $\ell_1$ error guarantee that $|\tilde{x}_i - x_i| = O(\| \mathbf{x} \|_1 / s)$ with constant probability (Chapter~3.5 in \cite{cormode2020book}).  Note that these two error bounds are in general incomparable. The success probability can be amplified to $1 - \beta$ via a standard median trick: creating $O(\log (1/\beta))$ independent instances and returning the median of the estimators. This way, a count sketch can be viewed as a matrix of $O(\log (1/\beta))$ rows and $s$ columns.  



\ifthenelse{\boolean{long}}{\subsection{Concentration Bounds}

We will employ a number of concentration bounds in our analysis (in addition to the standard Chernoff inequality), listed below in the order of increasing power but also requiring more conditions on the constituent random variables.

\begin{lemma}[Chernoff Bound]
Given independent Bernoulli random variables $Y_1, \ldots, Y_k$, such that $Y = \sum_{i=1}^k Y_i$ and $\mathsf{E}[Y] = \mu$, then, for $0 < \rho_1 \le 1$, and $0 \le \rho_2 \le 4$,
\begin{align*}
    &\Pr[Y \le (1-\rho_1)\mu] \le \exp\left(-\frac{\rho_1^2 \mu}{2}\right) \\
    &\Pr[Y \ge (1+\rho_2)\mu] \le \exp\left(-\frac{\rho_2^2 \mu}{4}\right) \\
\end{align*}
\end{lemma}

\begin{lemma}[Hoeffding Inequality]
Let $Y_1, \ldots, Y_k$ be independent random variables such that $|Y_i| \le K_i$ for all $i$. Then, for every $d \ge 0$, we have
\begin{align*}
\Pr\left[\left |\sum_{i=1}^k Y_i - \mathsf{E}\left[\sum_{i=1}^k Y_i\right] \right| \ge d \right] \le 2 \exp\left( -\frac{2 d^2}{\sum_{i=1}^{k} K_i^2} \right).
\end{align*}
\end{lemma}

\begin{lemma}[Bernstein Inequality]
Let $Y_1, \ldots, Y_k$ be independent, mean zero random variables such that $|Y_i| \le K$ for all $i$. Then, for every $d \ge 0$, we have
\begin{align*}
\Pr\left[\left |\sum_{i=1}^k Y_i\right| \ge d \right] \le 2 \exp\left( -\frac{d^2/2}{\sigma^2 + Kd/3} \right),
\end{align*}
where $\sigma^2 = \sum_{i=1}^k \mathsf{E}[Y_i^2]$ is the variance of the sum.
\end{lemma}

\begin{lemma} [Measure Concentration for Geometric Distribution \cite{chan2012optimal}]
Let $\alpha > 1$. Suppose $\{ Y_i \}_{i \in [k]}$ are $k$ independent random variables drawn from $\text{Geom}(\alpha)$, and let $Y = \sum_{i=1}^k Y_i$. Then
\begin{enumerate}
\item $\Pr\left[|Y| \ge \frac{4\alpha}{\alpha-1}\sqrt{k}\log \frac{2}{\beta}\right] \le \beta$;
\item If $k \ge \alpha \log \frac{2}{\beta}$, $\Pr\left[ |Y| \ge \frac{4\sqrt{\alpha}}{\alpha-1} \cdot \sqrt{k \log \frac{2}{\beta}} \right] \le \beta$.
\end{enumerate}
\end{lemma}

}{}

\section{One-shot Frequency Estimation}
\label{sec:one-shot}

In this section, we build up our approach. 
The starting point is a relatively simple protocol based on gathering a carefully configured sketch from each party. 
This is sufficient to give an accurate result (Section~\ref{sec:one-shot-base}). 
However, sending a large sketch can be costly when sites have few items, so we show how to reduce the sketch size for the LDP ($n_i=1$) case in Section~\ref{sec:ldp}, and to achieve a better communication cost in the case of variable input sizes in Section~\ref{sec:one-shot-hh}.

\subsection{Our Basic Protocol}
\label{sec:one-shot-base}

In addition to the privacy parameter $\varepsilon$ and the failure probability $\beta$, our protocol uses a parameter $s \ge 1$, which determines the average message size of each party.

\paragraph{Algorithm on Each Party.} On each party $i$, from the local counts $\mathbf{x}_i$ we build a count sketch $C_i$ of $R$ rows and $s_i := \lceil ks \cdot n_i / N \rceil$ columns, where $R$ is the nearest odd number from $\log \frac{3k}{\beta}$. Next, the party perturbs each counter $C_i[r, c]$, $r \in [R], c \in [s_i]$ in the sketch by adding noise $\eta_{r, c}^{(i)}$ drawn from $\text{Geom}(e^{\varepsilon/(2R)})$, to preserve privacy. Then the party sends this noisy count sketch $\Tilde{C}_i$ to the aggregator.  The communication cost (the total size of all count sketches) is
\begin{align*}
\textstyle
\sum_{i=1}^k s_i R = O\left(\sum_{i=1}^k ks \cdot \frac{n_i}{N} \log \frac{k}{\beta}\right) 
= O\left(ks \log\frac{k}{\beta}\right).
\end{align*}
If there is no public randomness, each party $i$ also needs to send the hash functions used in the count sketch $h_r^{(i)},g_r^{(i)},$ $r \in [R]$ to the aggregator, which takes $O(kR) = O(k\log\frac{k}{\beta})$ communication. 

\paragraph{Privacy Guarantee.} It is clear that a count sketch of $R$ rows has a sensitivity of $2R$, so adding noise drawn from $\text{Geom}(e^{\varepsilon/(2R)})$ is sufficient to preserve $\varepsilon$-DP for each party.

\paragraph{Algorithm on Aggregator.} After the aggregator has collected the noisy count sketch $\Tilde{C}_i$ from each party, for any $j \in [u]$, we use $\tilde{y}_j := \sum_i \operatorname{median}_{r \in [R]} \{ g_r^{(i)}(j) \cdot  \Tilde{C}_i[r, h_r^{(i)}(j)] \}$ as the estimator for $y_j := \sum_i x_{i,j}$.   Note that the way we combine the sketches is different from their standard use in the MDP/LDP model.  The common practice (e.g., \cite{bassily2017practical}) is to use the mergeability property of linear sketches, i.e., the aggregator merges the noisy sketches (so all parties must use the same hash functions) and makes the estimate from the merged sketch.  Instead, we make a separate estimate from each noisy sketch and add up the estimates.  Thus the parties do not use the same hash functions; actually, as shown in our analysis below, it is critical for the parties to use independent hash functions.  

\paragraph{Accuracy.} We use a lemma from \cite{wang2013quantiles},
\begin{lemma} [\cite{wang2013quantiles}]
\label{lem:med}
If $\{ X_i \}_{i \in [n]}$ are independent random variables, each of which has a symmetric PDF around zero, and $n$ is an odd number, then $\mathsf{E}\left[\operatorname{median}_{i \in [n]} \{ X_i \}\right] = 0$.
\end{lemma}
To see that $\mathsf{E}[\tilde{y}_j] = y_j$, first we show that the frequency estimator from each row of $\Tilde{C}_i$ is unbiased. For any $r \in [R]$,
\begin{align*}
&\mathsf{E}\left[g_r^{(i)}(j) \cdot  \Tilde{C}_i[r, h_r^{(i)}(j)] - x_{i,j}\right] = \mathsf{E}\left[g_r^{(i)}(j) \cdot C_i[r, h_r^{(i)}(j)] - x_{i,j}\right] + \mathsf{E}\left[g_r^{(i)}(j) \cdot \eta_{r, h_r^{(i)}(j)}^{(i)}\right] = 0.
\end{align*}
Moreover, since $g_r^{(i)}$ is a truly random hash function which maps $j$ to $\pm 1$ with equal probability, each random variable $g_r^{(i)}(j) \cdot \Tilde{C}_i[r, h_r^{(i)}(j)] - x_{i,j}$ has a symmetric PDF around zero.
Then, by Lemma~\ref{lem:med}, we have
\begin{align*}
&\mathsf{E}\left[\tilde{y}_j - y_j\right] = \sum_i \mathsf{E}\left[ \text{median}_{r \in [R]} \{ g_r^{(i)}(j) \cdot  \Tilde{C}_i[r, h_r^{(i)}(j)] - x_{i,j} \} \right] = 0.
\end{align*}
Next, we analyze the error $| \tilde{y}_j - y_j |$. Due to the $\ell_1$ error guarantee of the count sketch and the Chebyshev inequality for the Geometric noise, for any $i \in [k]$ and $r \in [R]$, we have
\begin{align*}
&\left| g_r^{(i)}(j) \cdot \Tilde{C}_i[r, h_r^{(i)}(j)] - x_{i,j} \right| \le \left| g_r^{(i)}(j) \cdot C_i[r, h_r^{(i)}(j)] - x_{i,j} \right| + \left| g_r^{(i)}(j) \cdot \eta_{r, h_r^{(i)}(j)}^{(i)} \right| \\
&= O\left(\frac{n_i}{s_i} + \frac{1}{e^{\varepsilon/(4R)}-1}\right) \qquad \text{with a constant probability, say, $0.95$}.
\end{align*}
Note that the frequency estimator from $\Tilde{C}_i$ is the median of the estimators from $R = \log\frac{3k}{\beta}$ independent rows.
We say that each estimate is good if it satisfies the above error bound, which happens with probability $0.95$, and let $Y$ be the number of estimates that are not good. The median estimator fails to be good with probability $\Pr[Y \ge \log\frac{3k}{\beta}/2] \le \exp(-81/80 \log\frac{3k}{\beta}) \le \beta/(3k)$ by the Chernoff bound. Thus, the success probability of the above error guarantee is amplified to $1 - \beta/(3k)$. 
Applying a union bound, this error guarantee holds for every party $i \in [k]$ with probability $1 - \beta/3$ --- let $E_1$ denote this event. 
Conditioned upon the event $E_1$, and since across $k$ parties the random variables $\text{median}_{r \in [R]} \{ g_r^{(i)}(j) \cdot  \Tilde{C}_i[r, h_r^{(i)}(j)] - x_{i,j} \}$ are independent and bounded by $O(N / (ks) + 1/(e^{\varepsilon/(4R)}-1))$, applying a Hoeffding bound we conclude that $| \tilde{y}_j - y_j | = O( N\sqrt{\log\frac{1}{\beta}}/(\sqrt{k}s) + \sqrt{k\log\frac{1}{\beta}} / (e^{\varepsilon/(4R)}-1) )$ with probability $1 - \beta/3$. Finally by the law of total probability, this error guarantee holds unconditionally with probability at least $1 - \beta$.

\begin{theorem}
\label{the:cs-one-shot}
For $s \ge 1$ and $\varepsilon > 0$, our $\varepsilon$-MDP one-shot frequency estimation protocol returns an unbiased estimator for the frequency of any item that with probability at least $1-\beta$ has error 
\[O\left( \left(\frac{N}{\sqrt{k}s} + \frac{\sqrt{k}}{e^{\varepsilon/(4\log\frac{3k}{\beta})}-1} \right) \cdot \sqrt{\log\frac{1}{\beta}}  \right).\]
Its expected communication cost is $O\left( ks \log\frac{k}{\beta} \right)$.
\end{theorem}

\subsection{A Refined Analysis for LDP}
\label{sec:ldp}

For LDP, which is a special case of MDP, by a more refined analysis based on the $\ell_2$ bound of the count sketch, we show that it is sufficient for each party to construct a count sketch of only one row and $s = \lceil (e^{\varepsilon/2}-1)^2/e^{\varepsilon/2} \rceil$ columns, and only the non-zero entries of the noisy sketch need to be sent to the aggregator.  In addition, it suffices for $h$ and $g$ to be both pairwise-independent hash functions.

\begin{theorem}
For $\varepsilon > 0$, our $\varepsilon$-LDP frequency estimation protocol returns an unbiased frequency estimator for any item with a variance of $O(k e^{\varepsilon/2} / (e^{\varepsilon/2}-1)^2)$, or an error of $O\left(\max \left\{ \sqrt{k\log\frac{1}{\beta}}/(e^{\varepsilon/4}-1), \log\frac{1}{\beta} \right\} \right)$ with probability $1 - \beta$. It uses $O(k \cdot \max\{ \varepsilon, \log \frac{1}{\varepsilon} \})$ bits of communication in expectation and $O(k \log u)$ bits of public randomness.
\end{theorem}

\ifthenelse{\boolean{long}}{
\begin{proof}
Let $s = \lceil (e^{\varepsilon/2}-1)^2/e^{\varepsilon/2} \rceil$ denote the number of columns of the count sketch, and let $g_i(j) \cdot \Tilde{C}_i[h_i(j)] = g_i(j) \cdot (C_i[h_i(j)] + \zeta_{i, h_i(j)})$ be the frequency estimation for an item $j$ from the noisy count sketch, where $\zeta_{i, h_i(j)}$ is drawn from Geom($e^{\varepsilon/2}$).
Each party needs to share two hash functions $g_i : [u] \rightarrow \{ -1, +1 \}$ and $h_i : [u] \rightarrow [s]$, which are used in count sketch, with the aggregator. Since these hash functions are pairwise-independent, each of them takes $O(\log u)$ bits of public randomness.

Now we analyze the error $|\sum_i g_i(j) \cdot \Tilde{C}_i[h_i(j)] - \sum_i x_{i, j} |$, which can be decomposed into two parts $| \sum_i g_i(j) \cdot C_i[h_i(j)] - \sum_i x_{i, j} |$ and $|\sum_i g_i(j) \cdot \zeta_{i, h_i(j)}|$. From the $L_2$ bound of count sketch, we know that $\mathsf{Var}[g_i(j) \cdot C_i[h_i(j)]] = O(1/s)$. Moreover, the absolute value of each $g_i(j) \cdot C_i[h_i(j)]$ is bounded by 1, so by the Bernstein inequality, we have $| \sum_i g_i(j) \cdot C_i[h_i(j)] - \sum_i x_{i, j} | = O\left(\sqrt{k\log\frac{1}{\beta} / s}\right)$ for $s = O(k/\log\frac{1}{\beta})$ with probability $1-\beta/2$. Note that the sign hash function $g_i$ will not change the distribution of $\zeta_{i, h_i(j)}$ which is symmetric around 0, so by the concentration property of Geometric distribution, we have $|\sum_i g_i(j) \cdot \zeta_{i, h_i(j)}| = O\left(\sqrt{k\log\frac{1}{\beta}}/(e^{\varepsilon/4}-1)\right)$ for $k \ge e^{\varepsilon/2} \log\frac{1}{\beta}$ with probability $1-\beta/2$. 
Finally, applying a union bound, the error is $O\left(\sqrt{k\log\frac{1}{\beta}}/(e^{\varepsilon/4}-1)\right)$ for $k \ge e^{\varepsilon/2} \log\frac{1}{\beta}$ with probability $1-\beta$. When $k < e^{\varepsilon/2} \log\frac{1}{\beta}$, we can always pad $k$ to $e^{\varepsilon/2} \log\frac{1}{\beta}$ by adding dummy parties, then by the above bound, the error in this case is $O(\log\frac{1}{\beta})$ with probability $1-\beta$. We conclude that the error is $O\left(\max \left\{ \sqrt{k\log\frac{1}{\beta}}/(e^{\varepsilon/4}-1), \log\frac{1}{\beta} \right\} \right)$ for any $k$ with probability $1 - \beta$. And it is easy to see that $\mathsf{Var}[\sum_i g_i(j) \cdot \tilde{C}_i[h_i(j)]] = \mathsf{Var}[\sum_i g_i(j) \cdot C_i[h_i(j)]] + \mathsf{Var}[\sum_i g_i(j) \cdot \zeta_{i, h_i(j)}] = O(k e^{\varepsilon/2} / (e^{\varepsilon/2}-1)^2)$.

Next, we analyze the communication cost. First, suppose $\varepsilon > 1$. Note that only the nonzero noisy counters need to be sent to the aggregator (both the value and the index), and the number of them is at most $O(1 + s/(1+e^{\varepsilon/2})) = O(1)$ in expectation, since the $\text{Geom}(e^{\varepsilon/2})$ noise takes nonzero value as output with probability $2/(1+e^{\varepsilon/2})$. So the communication cost (word complexity) is $O(1)$. To obtain bit complexity, note that each index takes $\log_2 s = O(\varepsilon)$ bits. And the value of each nonzero noisy counter can be encoded with $O(\varepsilon)$ bits in expectation. To prove this, let the random variable $X$ denote the value of a nonzero noisy counter, and note that $X = x + \eta'$, where $x$ is a counter in the original count sketch which can be $0$ or $\pm 1$, and $\eta'$ is drawn from Geom($e^{\varepsilon/2}$) conditioned upon the event that $\eta' \ne x$. Now we calculate an upper bound of $\mathsf{E}[|\eta'|]$. Let $\eta$ denote the random variable drawn from $\text{Geom}(e^{\varepsilon/2})$. We have $\mathsf{E}[|\eta|] \le \sqrt{\mathsf{E}[\eta^2]} = O(e^{\varepsilon/4}/(e^{\varepsilon/2}-1))$, and then by a standard calculation,
\begin{align*}
\mathsf{E}[|\eta'|] &= (\mathsf{E}[|\eta|] - p(\eta = x) \cdot |x|) / (1 - p(\eta = x)) \\
&\le \mathsf{E}[|\eta|] / (1 - p(\eta = 0)) \\
&= O((e^{\varepsilon/2}+1)/(e^{\varepsilon/2}-1) \cdot e^{\varepsilon/4}) \\
&= O(e^{\varepsilon/4}).
\end{align*}
Using Elias coding, the number of bits (in expectation) to encode $|X|$ is $\mathsf{E}\left[\log |X|\right] \le \log \mathsf{E}\left[|X|\right] \le \log \mathsf{E}[1 + |\eta'|] = O(\varepsilon)$ by Jensen's inequality. For the case $\varepsilon < 1$, we have $s = O(1)$. Then we simply send all noisy counters to the aggregator. The value of each noisy counter can be encoded with $O(\log \frac{1}{\varepsilon})$ bits, since $\mathsf{E}[|\eta|] = O(1/\varepsilon)$ for $\varepsilon < 1$.
\end{proof}}{The proof of the theorem and some others are presented in the supplementary material.}

\paragraph{Concurrent work.} In an independent and concurrent work \cite{feldman2021lossless}, Feldman and Talwar provide a general approach that compresses a LDP protocol with negligible loss in privacy and utility guarantees under standard cryptographic assumptions. When applied to frequency estimation, their protocol achieves error $O(\sqrt{k}/(e^{\varepsilon/4}-1))$ with communication cost $O(k \max\{ \log u, \varepsilon, \log\frac{1}{\varepsilon} \})$. If public randomness is available, our communication cost $O(k\max\{ \varepsilon, \log\frac{1}{\varepsilon} \})$ is less than theirs. 

\paragraph{Comparison to FreqOracle.} The frequency oracle used in \cite{bassily2017practical} combines the idea of count-sketch with the HRR protocol. It uses a Hadamard matrix of size $s'$, where $s' = \tilde{O}(\sqrt{k})$, and two hash functions $h : [u] \rightarrow [s'], g : [u] \rightarrow \{ -1, +1 \}$. Each party sends one bit $g(v_i) \cdot H[r_i, h(v_i)]$ via randomized response to the aggregator, where $r_i$ is uniformly drawn from $[s']$. The error of FreqOracle is the same as HRR asymptotically and restricted to the case $\varepsilon < 1$ as well\footnote{More precisely, the error is $O(\sqrt{k}\cdot (e^\varepsilon + 1)/(e^\varepsilon - 1)) = O(\sqrt{k}/\min \{ \varepsilon, 1 \})$. The reason is that in FreqOracle or HRR each party uniformly selects an entry of the
Hadamard matrix (for reducing communication cost), so even if $\varepsilon \rightarrow \infty$ the error is still $O(\sqrt{k})$. Chen et al.~\cite{chen2020breaking} improve the error to $O(\sqrt{k}/\sqrt{\varepsilon})$ for $\varepsilon > 1$, by using multiple samples from the Hadamard matrix to reduce the variance.}; their use of the count-sketch is to reduce the running time of identifying the heavy hitters from $\tilde{O}(k^{1.5})$ to $\tilde{O}(k)$. Our LDP protocol differs from FreqOracle in the following aspects: (1) we do not use a Hadamard matrix; (2) we set the width of the count-sketch as $s = \lceil (e^{\varepsilon/2}-1)^2/e^{\varepsilon/2} \rceil$ (regardless of $k$), while theirs is $s' = \tilde{O}(\sqrt{k})$ (regardless of $\varepsilon$); (3) we combine the count sketches in a different manner from \cite{bassily2017practical}.
On the other hand, the use of geometric noise instead of randomized response is not crucial: Both can be used to achieve error $1/(e^{\Theta(\varepsilon)}-1)$) in the LDP model, although geometric noise is needed in the MDP model where the counters in the sketch are not just $0$ and $1$. 

\ifthenelse{\boolean{long}}{
\paragraph{Comparison to OLH.} 
The OLH algorithm \cite{wang2017locally} improves over HRR for the case $\varepsilon>1$, although asymptotically it has the same error bound as HRR, i.e., the error is $\Omega(\sqrt{k})$ for $\varepsilon>1$ \footnote{Note that Wang et al.~\cite{wang2017locally} focus on $\mathsf{Var}^*$, which is only one part of the true variance (see Equation (3) in their paper).  
While $\mathsf{Var}^*$ decreases exponentially fast in $\varepsilon$, the full variance does not.}.
The first step of OLH is similar to ours, where each party hashes the item to $[s]$ for $s = e^\varepsilon + 1$, but without a sign hash function.  However, the crucial difference is the perturbation step. OLH perturbs the resulting hash value over $[s]$ using randomized response, i.e., reporting the true value with probability $e^\varepsilon/(e^\varepsilon + s - 1) = 1/2$, otherwise a value u.a.r.\ over $[s]$.  In our algorithm, we treat the hash value as the location in the count sketch.  We perturb the counters but not the location, which incurs much smaller errors when $\varepsilon>1$.}{}

\ifthenelse{\boolean{long}}{}{In the supplementary material, we also compare with OLH \cite{wang2017locally}.}

\paragraph{A Lower Bound.} 
To complement our upper bound, we prove a lower bound for $\varepsilon = O(\log k)$ by a reduction to the \textit{1-bit sum} problem under LDP, where each party holds a bit, and the aggregator wishes to estimate the number of $1$'s.

\begin{theorem}
Any LDP protocol for the 1-bit sum problem must have an error of $\Omega(\sqrt{k} / e^{\varepsilon/2})$ for $k = \Omega(e^\varepsilon)$.
\end{theorem}

\ifthenelse{\boolean{long}}{\begin{proof}
We need an anti-concentration bound from \cite{matouvsek2001probabilistic},

\begin{lemma}[\cite{matouvsek2001probabilistic}]
Let $Y$ be a sum of independent random variables, each attaining values in $[0, 1]$, and let $\sigma = \sqrt{\mathsf{Var}[Y]} \ge 200$. Then for all $0 \le d \le \sigma^2/100$, we have
\begin{align*}
\Pr[Y \ge \mathsf{E}[Y]+d] \ge c e^{-d^2/3\sigma^2}
\end{align*}
for a sufficiently small constant $c$.
\label{lem:anti}
\end{lemma}

We follow the same framework as \cite{chan2012optimal}. More precisely, consider a uniformly random input $\{ X_1, X_2, \ldots, X_k \} \in \{ 0, 1 \}^k$, from the LDP constraint it can be shown that for any possible transcript $\pi$,
\[ 1/(1+e^\varepsilon) \le \Pr[X_i=1 \mid \Pi=\pi] \le e^\varepsilon/(1+e^\varepsilon). \]
Then by Lemma~\ref{lem:anti}, we can obtain a lower bound of $\Omega(\mathsf{Var}[\sum_i X_i]) = \Omega(\sqrt{k e^{\varepsilon}}/(e^{\varepsilon}+1))$ for $k = \Omega(e^\varepsilon)$.
\end{proof}}{}

Note that the lower bound on the 1-bit sum problem also holds for the frequency estimation problem ($u \ge 2$), since any frequency estimation protocol can be used to solve the 1-bit sum problem. For $\varepsilon = O(\log k)$, Ghazi et al.~\cite{ghazi2019power} prove a lower bound of $\tilde{\Omega}(k^{1/3})$ (Equation~(11) in Theorem 3.3) that is independent of $\varepsilon$. In contrast, our lower bound characterizes the dependency on $\varepsilon$ and can be much better than theirs, for example, when $\varepsilon = \frac{1}{4} \log k$ ours is $\Omega(k^{3/8}) \gg \tilde{\Omega}(k^{1/3})$.
Combined with the lower bound of $\Omega(\sqrt{k}/\varepsilon)$ for $\varepsilon=O(1)$ in \cite{bassily2015local}, this implies our LDP protocol is optimal for $\varepsilon = O(\log k)$, up to a constant-factor difference in $\varepsilon$.

\subsection{Further Improvement by Frequency Separation}
\label{sec:one-shot-hh}
In the protocol described in Section~\ref{sec:one-shot-base}, the sketch size $s_i$ is proportional to $n_i$ on each party, even if it has only one item with local count $n_i$. This results in a large number of informationless noisy counters to be sent to the aggregator. 
\ifthenelse{\boolean{long}}{In this subsection, we describe a method to reduce the communication cost, which works particularly well on skewed data while providing the same worst case guarantee as Theorem~\ref{the:cs-one-shot}.  
The idea is to divide the local counts into ``heavy'' and ``light'' groups.  
We use an importance sampling based method for the heavy items, while dealing with the rest using count sketch as before. We also demonstrate its effectiveness in the experiments.

More precisely, we separate the local counts $\{x_{i,j}\}_j$ into \textit{local heavy hitters} and \textit{local light hitters}.  To preserve privacy, we do so probabilistically, as follows: (1) draw a noise vector $\mathbf{\xi}_i \in \mathbb{R}^u$, where each coordinate is i.i.d. from $\text{Geom}(e^{\varepsilon/4})$, and perturb $\mathbf{x}_i$ as $\mathbf{\tilde{x}}_i := \mathbf{x}_i + \mathbf{\xi}_i$; (2) extract entries $j \in [u]$ such that $\tilde{x}_{i,j} > T$ as the local heavy hitters, for some threshold $T = \Theta(\frac{1}{\varepsilon} \log (ku))$, while the others are the local light hitters. We denote the identities of the local heavy and light hitters at party $i$ as $S^{\mathrm{hi}}_i$ and $S^{\mathrm{lo}}_i$, respectively.  
All error analyses below hold for any fixed separation of the local heavy/light hitters, i.e., conditioned upon $S^{\mathrm{hi}}_i$ and $S^{\mathrm{lo}}_i$ for all $i \in [k]$; by the law of total probability the error guarantee will hold unconditionally.


In order to avoid running time proportional to $u$, the above procedure for separating $[u]$ into $S^{\mathrm{hi}}$ and $S^{\mathrm{lo}}$ can be equivalently done as follows,

\begin{enumerate}
\item Add i.i.d. noise drawn from $\text{Geom}(e^{\varepsilon/4})$ to the non-zero entries of $\mathbf{x}_i$, and extract the entries with noisy count above the given threshold $T$ as $S^{\mathrm{hi}}$.
\item For the zero entries of $\mathbf{x}_i$, first draw $m_i \sim \text{Binomial}(u - n_i, p_{T})$, where $p_{T} = \exp((-T+1)\varepsilon/4)/(\exp(\varepsilon/4)+1)$ is the probability that a zero entry has a noisy count above $T$. Then, uniformly at random select $m_i$ locations from the zero entries of $\mathbf{x}_i$ and add them to $S^{\mathrm{hi}}$.
\end{enumerate}

It is safe for each party to release $S^{\mathrm{hi}}_i$ and $S^{\mathrm{lo}}_i$.  This is because the local counts $\{x_{i,j}\}_j$  have a sensitivity of $2$, thus adding noise drawn from $\text{Geom}(e^{\varepsilon/4})$ is sufficient to preserve $\varepsilon/2$-DP. Then by the post-processing property of DP, the separation results are $\varepsilon/2$-DP. 
In the following, we present DP mechanisms for the local heavy and light hitters respectively.  The input to these mechanisms are the identities of the local heavy/light hitters and their (original) local counts.  Since the locations of the local heavy and light hitters are disjoint, by the parallel composition theorem, it suffices for these two mechanisms to be $\varepsilon/2$-DP.

\subsubsection{Local Heavy Hitters}
\paragraph{Algorithm on Each Party.} For each $j \in S^{\mathrm{hi}}_i$ at party $i$, we perturb $x_{i,j}$, with fresh noise $\zeta_{i, j}$ drawn from $\text{Geom}(e^{\varepsilon'/2})$, as $\hat{x}_{i,j} := x_{i, j} + \zeta_{i, j}$, where $\varepsilon' = \varepsilon/\log\frac{1}{\beta}$.  
Still, sending all such items would consume a lot of communication so instead we send these perturbed local counts by importance sampling.  More precisely, party $i$ sends each pair $(j, \hat{x}_{i,j}/p), j\in S_i^{\mathrm{hi}}$ to the aggregator with probability $p(|\hat{x}_{i,j}|)$, where $p(x) = \min \{ ks \cdot x / N, 1 \}$. This procedure (perturb and importance sampling) is repeated by $\frac{1}{2}\log\frac{1}{\beta}$ times.

\paragraph{Privacy Guarantee.} Because the frequency vector has a sensitivity of 2, adding noise drawn from $\text{Geom}(e^{\varepsilon'/2})$ provides the guarantee of $\varepsilon'$-DP. By the basic composition theorem across all repetitions, the whole procedure preserves $\varepsilon/2$-DP.

\paragraph{Communication.} The expected communication cost in each repetition is
\begin{align*}
&
\mathsf{E}\left[\sum_{i=1}^k \sum_{j \in S^{\mathrm{hi}}_i} p(\hat{x}_{i,j})\right] = \mathsf{E}\left[ \mathsf{E}\left[\sum_{i=1}^k \sum_{j \in S^{\mathrm{hi}}_i} p(\hat{x}_{i,j})  \,\middle|\,  S^{\mathrm{hi}} \right] \right] \\
&\le \mathsf{E}\left[ \mathsf{E}\left[\sum_{i=1}^k \sum_{j \in S^{\mathrm{hi}}_i} ks \cdot |\hat{x}_{i,j}| / N  \,\middle|\,  S^{\mathrm{hi}} \right] \right] \\
&\le \frac{ks}{N} \cdot \mathsf{E}\left[ \mathsf{E}\left[\sum_{i=1}^k \sum_{j \in S^{\mathrm{hi}}_i} x_{i,j} + |\zeta_{i, j}|  \,\middle|\,  S^{\mathrm{hi}} \right] \right] \\
&= \frac{ks}{N} \cdot \mathsf{E}\left[ \sum_{i=1}^k \sum_{j \in S^{\mathrm{hi}}_i} x_{i,j} + \sum_{i=1}^k O\left(\frac{1}{\varepsilon}\right) \cdot |S^{\mathrm{hi}}_i| \right] \\
&\le ks + \frac{ks}{N} \cdot O\left(\frac{1}{\varepsilon}\right) \cdot \mathsf{E}\left[ \sum_{i=1}^k |S^{\mathrm{hi}}_i| \right].
\end{align*}

To bound $\mathsf{E}\left[ \sum_{i=1}^k |S^{\mathrm{hi}}_i| \right]$, let $E$ denote the event that $| \xi_{i, j} | = O(\frac{1}{\varepsilon} \log (ku))$ simultaneously for every $i \in [k], j \in [u]$.
Then we have $\Pr[E] \ge 1- 1/(ku)$ by the tail property of the Geometric distribution and applying a union bound. Furthermore, conditioned upon $E$, every local heavy hitter has a frequency of $\Omega( \frac{1}{\varepsilon} \log(ku))$. Then, we have
\begin{align*}
\mathsf{E}\left[ \sum_{i=1}^k |S^{\mathrm{hi}}_i| \right]  &\le  \Pr[E] \cdot \mathsf{E}\left[ \sum_{i=1}^k |S^{\mathrm{hi}}_i|  \,\middle|\,  E \right] + \Pr[\Bar{E}] \cdot ku   \\
&\le O\left( \frac{N}{\frac{1}{\varepsilon} \log(ku)} \right).
\end{align*}
Thus expected communication cost is $O(ks\log\frac{1}{\beta})$ across all repetitions.

\paragraph{Algorithm on Aggregator.} In each repetition, let $g_{i,j}$ denote the HT estimator for $\hat{x}_{i,j}$. More precisely, if the aggregator received $\hat{x}_{i,j}$, we use $g_{i,j} = \hat{x}_{i,j} / p(|\hat{x}_{i,j}|)$, otherwise $g_{i,j} = 0$. The aggregator uses $\tilde{y}_j^{\mathrm{hi}} := \sum_i g_{i,j}$ as the estimate for $y_j^{\mathrm{hi}} := \sum_{i : j \in S^{\mathrm{hi}}_i} x_{i,j}$ in each repetition, and takes the median of these estimates across all $\frac{1}{2}\log\frac{1}{\beta}$ repetitions as the final estimator.

\paragraph{Accuracy.} It suffices to show that the estimate in each repetition satisfies the desired error guarantee with a constant probability, say, $0.99$, then the success probability of the median estimate from all $\frac{1}{2}\log\frac{1}{\beta}$ repetitions can be amplified to $1-\beta$ by the Chernoff bound. First, we show that $\tilde{y}_j^{\mathrm{hi}}$ is an unbiased estimator of $y_j^{\mathrm{hi}}$, let $\hat{x}$ denote $\{ \hat{x}_{i,j} \}_{i : j \in S^{\mathrm{hi}}_i}$,
\begin{align*}
\mathsf{E}[\tilde{y}_j^{\mathrm{hi}}] = \mathsf{E}\left[ \mathsf{E}\left[ \sum_i g_{i,j} \,\middle|\, \hat{x} \right] \right] 
= \mathsf{E}\left[ \sum_{i : j \in S^{\mathrm{hi}}_i} \hat{x}_{i,j}\right] 
= \sum_{i : j \in S^{\mathrm{hi}}_i} x_{i,j}.
\end{align*}
Next, we analyze the error $| \tilde{y}_j^{\mathrm{hi}} - y_j^{\mathrm{hi}} |$, which is composed of two parts: (1) $\sum_{i} g_{i,j}$ approximates $\sum_{i : j \in S^{\mathrm{hi}}_i} \hat{x}_{i,j}$; (2) $\sum_{i : j \in S^{\mathrm{hi}}_i} \hat{x}_{i,j}$ approximates $\sum_{i : j \in S^{\mathrm{hi}}_i} x_{i,j}$.

For part (1), let $\beta' = 0.01$, we first show that the error is $O(N/(\sqrt{k} s))$ with probability at least $1 - \beta'/2$ for any fixed choice of $\hat{x}$, i.e., conditioned upon the randomness of $\hat{x}$, then by the law of total probability the same error guarantee holds unconditionally. 
It suffices to consider the worst case that for all $i \in [k]$, $p(|\hat{x}_{i, j}|) < 1$, otherwise $g_{i,j}=\hat{x}_{i,j}$ which is already correct. 
Let $F$ denote the event that $| \sum_{i} g_{i,j} - \sum_{i : j \in S^{\mathrm{hi}}_i} \hat{x}_{i,j} | = O( N/(\sqrt{k} s) )$. Since $\mathsf{E}[\sum_i g_{i,j} \mid \hat{x}] = \sum_{i : j \in S^{\mathrm{hi}}_i} \hat{x}_{i,j}$ and $|g_{i,j}| \le N / (ks)$, by a Hoeffding bound we have $\Pr [F \mid \hat{x} ] \ge 1 - \beta'/2$.
Then, by the law of total probability,
\begin{align*}
\textstyle
\Pr [F] = \sum_{ \hat{x} } \Pr [F \mid \hat{x} ] \cdot p(\hat{x}) 
\ge \sum_{ \hat{x} } (1 - \frac{\beta'}{2}) \cdot p(\hat{x}) 
= 1 - \frac{\beta'}{2}.
\end{align*}
We conclude that the error for part (1) is $O(N/(\sqrt{k} s))$ with probability at least $1 - \beta'/2$.


For part (2), since each $\zeta_{i, j}$ is drawn from $\text{Geom}(e^{\varepsilon/(2\log\frac{1}{\beta})})$, by the Chebyshev inequality, it is easy to see that $|\sum_i \zeta_{i, j}| = O(\sqrt{k}/(e^{\varepsilon/(4\log\frac{1}{\beta})}-1))$ with probability at least $1 - \beta'/2$.

Finally, applying a union bound, the error $| \tilde{y}_j^{\mathrm{hi}} - y_j^{\mathrm{hi}} |$ is $O( N / (\sqrt{k} s) + \sqrt{k}/(e^{\varepsilon/(4\log\frac{1}{\beta})}-1) )$ with probability at least $1 - \beta' = 0.99$.

\subsubsection{Local Light Hitters}

We apply our count sketch based method (with privacy parameter $\varepsilon/4$ and $R = \frac{1}{4}\log\frac{3k}{\beta}$) over the local light hitters. Let $n_i^{\mathrm{lo}}$ denote the total (true) frequency of the local light hitters. Recall that the sketch size is proportional to $n_i^{\mathrm{lo}}$, which in this case is sensitive information and cannot be directly released. So we use $s_i := \lceil ks \cdot \tilde{n}_i^{\mathrm{lo}} / N \rceil$ as the sketch size, where $\tilde{n}_i^{\mathrm{lo}} = \min \{ n_i^{\mathrm{lo}} + \frac{8}{\varepsilon} \log \frac{2k}{\beta} + \text{Geom}(e^{\varepsilon/4}), n_i \}$ is an upper bound of $n_i^{\mathrm{lo}}$ for every $i$ with probability $1 - \beta/2$. Conditioned upon $\tilde{n}_i^{\mathrm{lo}} \ge n_i^{\mathrm{lo}}$, the error bound in Theorem~\ref{the:cs-one-shot} holds for local light hitters with probability $1-\beta/2$. Then by a union bound, this error guarantee holds unconditionally. Moreover, since $\tilde{n}_i^{\mathrm{lo}} \le n_i$, the communication bound in Theorem~\ref{the:cs-one-shot} also holds.

\paragraph{Privacy Guarantee} Note that $\| \mathbf{x}_i^{\mathrm{lo}} \|_1$ has a sensitivity of 1, then adding $\text{Geom}(e^{\varepsilon/4})$ noise to it is sufficient to preserve $\varepsilon/4$-DP. And our count sketch based method provides a guarantee of $\varepsilon/4$-DP. Then by the basic composition theorem, the protocol applied on the local light hitters preserves $\varepsilon/2$-DP.

}{In the supplementary material, we describe a method to reduce the communication cost, which works particularly well on skewed data while providing the same worst case guarantee as Theorem~\ref{the:cs-one-shot}.  
The idea is to divide the local counts into ``heavy'' and ``light'' groups.  
We use an importance sampling based method for the heavy items, while dealing with the rest using count sketch as before. We also demonstrate its effectiveness in the experiments.}

\section{Streaming Protocols}
In the streaming MDP model, each of the $k$ parties receives a stream of items, one at each time step.  Let $n$ be the total number of time steps.  For simplicity, we assume that $n$ is known to the protocol in advance; standard techniques can be used to remove this assumption, while incurring some extra logarithmic factors  in the error and costs \cite{chan2011private}.  Let $v_{i, t}$ denote the item received by party $i$ at time step $t$, and $f(v; t_1, t_2)$ the frequency of a given item $v$ received across all parties between time step $t_1$ and $t_2$ (inclusive).
\ifthenelse{\boolean{long}}{In Section \ref{sec:abserr} we first present a protocol that maintains a synopsis from which an estimate of $f(v; 1, t)$ can be extracted for any $v$ at each time step $t$; in Section \ref{sec:stream-sw} we extend it to the sliding-window model, i.e., we estimate $f(v; t-w+1, t)$ where $w$ is the window length.
}{In the following, we present a protocol that maintains a synopsis from which an estimate of $f(v; 1, t)$ can be extracted for any $v$ at each time step $t$; and in the supplementary material we extend it to the sliding-window model, i.e., we estimate $f(v; t-w+1, t)$ where $w$ is the window length.}


\subsection{Full-stream Protocol}
\label{sec:abserr}
Let $s \ge 1$ and $\Delta := \lceil n/s \rceil \cdot \sqrt{k}$.  We divide the stream into $m := \sqrt{k} \cdot n / \Delta = \min \{ n ,s \}$ epochs of $b := \Delta/\sqrt{k}$ time steps each. We say that an epoch is \textit{complete} if items in all time steps in this epoch have been received, otherwise we say that it is \textit{active}. Express the current time as $t = q \cdot b + r$ where $q, r \in \mathbb{Z}$ and $0 \le r < b$. To estimate $f(v; 1, t)$, we estimate $f(v; 1, qb)$ (i.e., over all complete epochs) and $f(v; qb+1, t)$ (i.e., over the current active epoch) separately using different methods.  The \textit{intra-epoch} protocol, which estimates $f(v; qb+1, t)$, operates on a per time step basis, so its error (variance) grows linearly as time goes on and would be too large beyond one epoch. The \textit{inter-epoch} protocol works on the epoch level; it responds slower to the stream but its error only grows logarithmically. 

\subsubsection{Intra-epoch protocol}

First of all, note that we only need to run the intra-epoch protocol when $b > 1$, or $s < n$.

\paragraph{Algorithm on Each Party.} 
Each party samples each time step $t$ with probability $p = b^{-1}=\sqrt{k}/\Delta$ independently. If $t$ is sampled, the party encodes the item $v_{i, t}$ as $\mathcal{M}(v_{i, t})$ using HRR (with privacy parameter $\varepsilon/2$) and sends it to the aggregator.

\paragraph{Algorithm on Aggregator.} The aggregator collects the messages received during the current active epoch and calculates
\begin{align*}
\textstyle
\tilde{f}(v; qb+1, t) = \frac{1}{p} \cdot \sum_{i} \sum_{t' \in S_i} \tilde{f}_{\mathcal{M}(v_{i, t'})}(v)
\end{align*}
as the estimation for $f(v; qb+1, t)$, where $S_i$ denotes the time steps sampled at each party $i$ during this epoch, and $\tilde{f}_{\mathcal{M}(v_{i, t'})}(v)$ denotes the frequency estimator for item $v$ used in the HRR protocol.

\paragraph{Communication.} As in each epoch of size $b$ each party samples an item with probability $p = b^{-1}$, the expected communication cost in each epoch is $O(k \cdot b\cdot b^{-1}) = O(k)$, and across all epochs the total is $O(km) = O(ks)$.

\ifthenelse{\boolean{long}}{\paragraph{Accuracy.} 
We first show that the estimator is unbiased. Let the random variable $Z_{i, t'} \in \{ 0, 1 \}$ indicate whether the time step $t'$ gets sampled, and let $X_{i, t'} := \tilde{f}_{\mathcal{M}(v_{i, t'})}(v)$ denote the frequency estimator used in the HRR protocol. Then the estimator can be written as
\begin{align*}
\textstyle
\tilde{f}(v; qb+1, t) = \frac{1}{p} \cdot \sum_i \sum_{qb < t' \le t} Z_{i, t'} X_{i, t'}.
\end{align*}
Note that we have $\mathsf{E}[X_{i, t'}] = x_{i, t'}$, where $x_{i, t'} \in \{ 0, 1 \}$ indicates whether $v_{i, t'}$ is identical to $v$.  By the independence of $Z_{i, t'}$ and $X_{i, t'}$, we conclude that $\tilde{f}(v; qb+1, t)$ is an unbiased estimator for $f(v; qb+1, t)$.

Since our algorithm essentially runs HRR over the sampled time steps, the error consists of two parts: the error due to sampling and the error due to HRR:
\begin{align*}
&|\tilde{f}(v; qb+1, t) - f(v; qb+1, t)| = \left|\frac{1}{p} \cdot \sum_i \sum_{qb < t' \le t} Z_{i, t'} X_{i, t'} - \sum_i \sum_{qb < t' \le t} x_{i, t'}\right| \\
&\le \left|\frac{1}{p} \cdot \sum_i \sum_{qb < t' \le t} Z_{i, t'} \cdot (X_{i, t'} - x_{i, t'})\right| + \left|\frac{1}{p} \cdot \sum_i \sum_{qb < t' \le t} Z_{i, t'} x_{i, t'} - \sum_i \sum_{qb < t' \le t} x_{i, t'}\right|.
\end{align*}

We first bound the first error term. 
Denote the number of items sampled during this epoch as $Z^{(qb : (q+1)b)} := \sum_i \sum_{qb < t' \le (q+1)b} Z_{i, t'}$. Observe that $\mathsf{E}[Z^{(qb : (q+1)b)}] = k \cdot b \cdot p = k$ and $\mathsf{Var}[Z^{(qb : (q+1)b)}] \le k$. 
Hence, by the Bernstein inequality, we have $Z^{(qb : (q+1)b)} = O(\max \{ k, \log\frac{1}{\beta} \})$ with probability at least $1 - \beta/3$. 
Conditioned upon this event and by the accuracy guarantee of HRR, with probability at least $1 - \beta/3$, the first error term is bounded by $O(\frac{1}{p} \sqrt{\max \{ k, \log\frac{1}{\beta} \} \log \frac{1}{\beta}} / \min \{ \varepsilon, 1 \}) = O(\Delta \log \frac{1}{\beta} / \min \{ \varepsilon, 1 \})$. 
Then the same error guarantee holds unconditionally with probability at least $1 - 2\beta/3$.

For the second error term, we first bound the variance
\begin{align*}
\mathsf{Var}\left[ \frac{1}{p} \cdot \sum_i \sum_{qb < t' \le t} Z_{i, t'} x_{i, t'} \right] &= \sum_i \sum_{qb < t' \le t} \mathsf{E} \left[ \left(\frac{1}{p} \cdot Z_{i, t'} x_{i, t'} - x_{i, t'} \right)^2\right] \\
&= \sum_i \sum_{qb < t' \le t} \frac{1-p}{p} x_{i, t'}^2 \\
&\le k \cdot \frac{\Delta}{\sqrt{k}} \cdot \frac{1-p}{p} \\
&\le \Delta^2.
\end{align*}
Note that each $|\frac{1}{p} \cdot Z_{i, t'} x_{i, t'} - x_{i, t'}|$ is bounded by $O\left(\frac{1}{p}\right) = O\left(\frac{\Delta}{\sqrt{k}}\right)$.  By the Bernstein inequality, we conclude that the error for the second part is $O(\Delta \log \frac{1}{\beta} )$ with probability at least $1 - \beta/3$.

Finally, applying a union bound, with probability at least $1 - \beta$, the error of the estimator $\tilde{f}(v; qb+1, t)$ is $O( \Delta \log \frac{1}{\beta} / \min \{ \varepsilon, 1 \} ) = O( n \sqrt{k} \log \frac{1}{\beta} / (s \cdot \min \{ \varepsilon, 1 \}) )$.


\paragraph{Privacy Guarantee.} Observe that the sampling procedure is data-independent and the information of each item is released only once. So the intra-epoch protocol guarantees $\varepsilon/2$-DP, as provided by the HRR protocol.

\paragraph{Space/time.} Observe that each party does not need store any historical information before time step $t$, so the intra-epoch protocol needs $O(1)$ space on each party.  Sampling and running HRR take $O(1)$ time per time step.}{}

\subsubsection{Inter-epoch protocol}

To obtain estimation for $f(v; 1, qb)$, we make use of a dyadic structure, which naturally corresponds to a tree representation, imposed over all epochs. Specifically, we build $\log m$ levels, and for each level $l$ the epoch's time steps are divided into $n / (2^l \cdot b)$ consecutive blocks of size $2^l \cdot b$ each. More precisely, for $0 \le l < \log m, 1 \le j \le n / (2^l \cdot b)$, let $B_{l, j} = \{ t \mid (j-1) \cdot 2^l \cdot b < t \le j \cdot 2^l \cdot b \}$ denote the $j$-th block at level $l$.  Note that each block on level 0 corresponds to an epoch.  Essentially, the inter-epoch protocol runs our one-shot algorithm for each block, but using different parameters.

\paragraph{Algorithm on Each Party.} Each party $i$ maintains a count sketch of $R = \frac{1}{2}\log \frac{2 k \log m}{\beta}$ rows and $s \sqrt{\log m} \cdot |B|/n$ columns for the items within each block $B$, where $l$ denotes the level of this block and $|B| = 2^l \cdot b$ denotes the number of time steps in $B$. After $B$ completes, we add i.i.d.\ noise draw from $\text{Geom}(e^{\varepsilon/(4R \log m)})$ to each counter in the count sketch, then send this noisy count sketch to the aggregator.

\paragraph{Algorithm on Aggregator.} As in our one-shot protocol, from the noisy count sketches (across all parties) corresponding to each block, the aggregator can obtain a frequency estimator for any item within this block. Furthermore, we know that the interval $[1, qb]$ can be decomposed into at most $\log m$ disjoint dyadic blocks, at most one from each level. Thus, to obtain an estimation for $f(v; 1, qb)$, we just add up the frequency estimates for $v$ from these blocks.

\paragraph{Communication.} The communication cost (the total size of all count sketches) for each block $B$ (across all parties) is $O(k R \cdot |B|/n \cdot s\sqrt{\log m})$, and the total (across all blocks) is $O(k R \cdot s\log^{1.5} m)$.

\ifthenelse{\boolean{long}}{\paragraph{Accuracy.} According to the analysis in Section~\ref{sec:one-shot}, from each noisy count sketch, the frequency estimation for any item is unbiased (by Lemma~\ref{lem:med}), and by a Chernoff bound across $R$ rows (where the constant failure probability for the estimate from each row is set to $0.01$), the error is $O(n/(s \sqrt{\log m}) + 1/(e^{\varepsilon/(8R\log m)}-1))$ with probability at least $1 - \beta/(2k\log m)$. Applying a union bound, across the $O(\log m)$ blocks on $k$ parties which constitutes the interval $[1, qb]$, this error guarantee holds simultaneously with probability at least $1 - \beta/2$. Then conditioned upon this event, we can apply a Hoeffding bound across $O(k \log m)$ frequency estimations from each noisy count sketch, and we conclude that the error is $O(n \sqrt{k\log\frac{1}{\beta}}/s + \sqrt{k \log m \log\frac{1}{\beta}}/(e^{\varepsilon/(8R\log m)}-1))$ with probability at least $1 - \beta/2$. Finally by the law of total probability, this error guarantee holds unconditionally with probability at least $1 - \beta$.

\paragraph{Privacy Guarantee.} Observe that each item contributes to $\log m$ blocks, so by the basic composition theorem, in order to provide a guarantee of $\varepsilon/2$-DP, it suffices to preserve $\varepsilon/(2\log m)$-DP for each block. It is clear that a count sketch of $R$ rows has a sensitivity of $2R$, thus adding noise drawn from $\text{Geom}(e^{\varepsilon/(4 R \log m)})$ preserves $\varepsilon/(2\log m)$-DP.

\paragraph{Space/time.} For each dyadic level, we only need keep the frequency sketch of the most recent complete block and the currently active block, since all previous blocks will no longer be useful. Then the space consumption on each party is $O(R \cdot s\sqrt{\log m}/n \cdot (n + n/2 + n/4 + \ldots)) = O(R \cdot s\sqrt{\log m})$. As each item contributes to $O(\log m)$ blocks and each block maintains a count sketch of $R$ rows, the time to process each item is $O(\log s \log\frac{k\log s}{\beta})$.}{}

Combining the intra-epoch and inter-epoch algorithm, we obtain the following result.

\begin{theorem}
\label{the:abs-err}
For $\varepsilon > 0$ and $s \ge 1$, our $\varepsilon$-MDP streaming frequency estimation protocol is able to return, at each time step, an unbiased estimator for the frequency of any item.   With probability at least $1-\beta$, the error of the estimator is
\[
O\left(\frac{n\sqrt{k}\log\frac{1}{\beta}}{s \cdot \min \{ \varepsilon, 1 \}} + \frac{\sqrt{k\log s\log\frac{1}{\beta}}}{e^{\varepsilon/(4\log s\log\frac{k\log s}{\beta})}-1}\right)
\]
for $s < n$, or
\[
O\left(\frac{n\sqrt{k\log\frac{1}{\beta}}}{s} + \frac{\sqrt{k\log n\log\frac{1}{\beta}}}{e^{\varepsilon/(4\log n\log\frac{k\log n}{\beta})}-1}\right)
\]
for $s \ge n$ (in this case the intra-epoch algorithm is not needed).
Its expected communication cost is $O(ks \cdot \log^{1.5} s \log\frac{k\log s}{\beta})$, and it takes $O(s \cdot \sqrt{\log s} \log\frac{k\log s}{\beta})$ space and $O(\log s \log\frac{k\log s}{\beta})$ time to process each item on each party.
\end{theorem}

\ifthenelse{\boolean{long}}{Note that the $s\ge n$ (or $s \ge w$ in the sliding-window model) case is meaningful only logarithmic factors are considered, which is why it does not show up in Figure \ref{fig:result}.}{}

The streaming MDP model degenerates into the continual observation model with $k:=1$.  In this case, each epoch has $b=\lceil n/s \rceil$ items, so we may just discard them, and only run the inter-epoch algorithm.  We then obtain a streaming algorithm with the following space-error trade-off:
\begin{corollary}
\label{cor:stream}
For $s \ge 1$, our $\varepsilon$-DP frequency estimation algorithm runs on a single stream using space $O(s \cdot \sqrt{\log s} \log\frac{\log s}{\beta})$.  At any time, an estimate for the frequency of any item can be extracted that, with probability $1-\beta$, has error 
$O\left(n\sqrt{\log\frac{1}{\beta}}/s + \sqrt{\log s\log\frac{1}{\beta}}/(e^{\varepsilon/(4\log s\log\frac{\log s}{\beta})}-1)\right)$.
\end{corollary}


\paragraph{Comparison with Chan et al.} There are two major differences between our protocol with Chan et al.~\cite{chan2012differentially}: 
(1) the choice of frequency summary technique (the MG algorithm v.s. count sketch); 
and (2) the introduction of an intra-epoch protocol. Intuitively, (1) is important because for a sketch size $s$, the MG algorithm incurs a bias of $O(\frac{w}{s})$ on each party, so the resulting protocol has an error proportional to $k$, in contrast to the factor-$\sqrt{k}$ in our error bound. Moreover, MG has sensitivity of $s$, so the error of their protocol is at least $\tilde{\Omega}(k\sqrt{w})$ (for any $s$ and a constant $\varepsilon$); while count sketch has sensitivity $1$, and our protocol can achieve an error $\tilde{O}(\sqrt{k})$.  (2) is important 
to achieve optimal error:
the protocol in \cite{chan2012differentially} doesn't need such an intra-epoch component, because their one-shot algorithm already has an error proportional to $k$, so they can simply ignore all items inside an epoch.  
We aim at the optimal error proportional to $\sqrt{k}$, which requires a more careful handling of the intra-epoch items.

\ifthenelse{\boolean{long}}{\subsection{Sliding-Window Protocol}
\label{sec:stream-sw}
Given a window length $w$, our full-stream protocol can be easily turned into a sliding-window protocol.  For simplicity we assume that $w$ is a multiple of $b$. We still run the intra-epoch algorithm for each epoch.  We run the inter-epoch algorithm on each \textit{tumbling window}, i.e., the time steps $(qw, (q+1)w]$ for each integer $q \ge 0$.  The only difference is that, in the full-stream protocol, we only keep the count sketches of the two most recent blocks on each dyadic level, while in the sliding-window case, we keep the sketches of all blocks in the two most recent tumbling windows.  This results in an $O(\log s)$ factor increase in space.

\begin{figure}[htbp]
    \centering
    \includegraphics[width=0.8\textwidth]{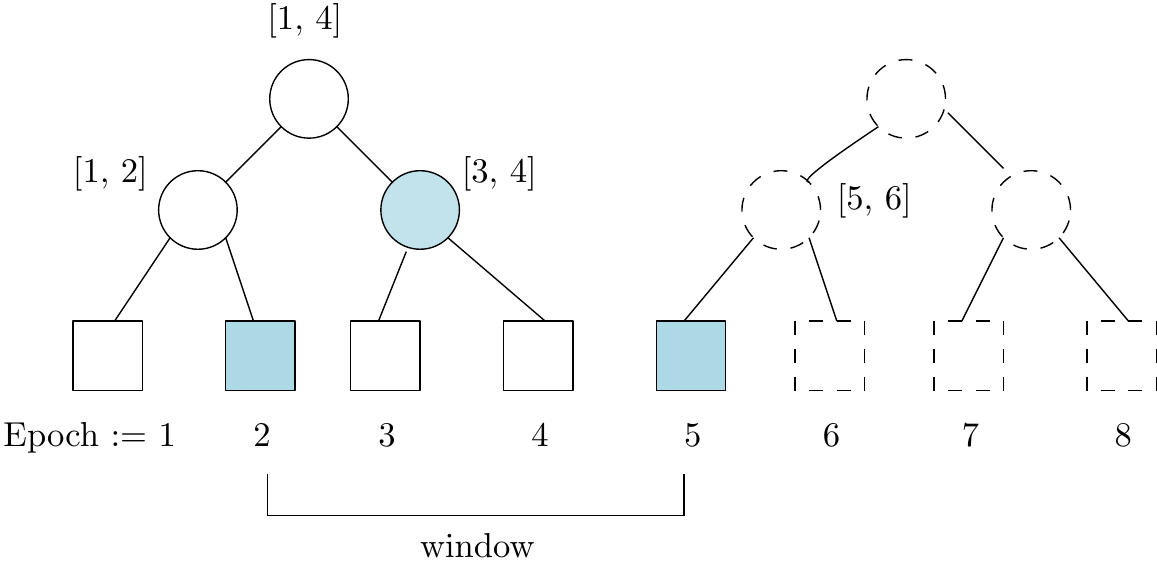}
     \caption{Tracking over a sliding window.}
     \label{fig:exa_sw}
\end{figure}

With all the blocks in the inter-epoch algorithm and the samples from the intra-epoch algorithm in the two most recent tumbling windows, the aggregator is able to extract a frequency estimate for any item at any time.  Figure~\ref{fig:exa_sw} shows an example where the window length is $w=4b$. Suppose the current time is in epoch $5$. The current sliding window can be decomposed into three parts: an intra-epoch part in epoch $2$, an intra-epoch part in epoch $5$, and an inter-epoch part covering epochs $3, 4$.  The frequency in the two intra-epoch parts can be estimated from the samples collected in these two epochs, while the inter-epoch part can be covered by at most $2\log m$ blocks in the two dyadic trees.  This is just a factor-$2$ difference compared with the full-stream case.  We thus conclude that Theorem \ref{the:abs-err} and Corollary \ref{cor:stream} continue to hold in the sliding-window model by simply replacing $n$ with $w$, except that (1) space will increase by an $O(\log s)$ factor, and (2) the communication cost is incurred per window.}{}

\section{Experiments}

\ifthenelse{\boolean{long}}{
\subsection{One-Shot}





We evaluate our method for finding heavy hitters using the following standard metrics (see e.g., \cite{cormode2010methods}):
(1) Recall, namely the number of true heavy hitters reported over the number of all true heavy hitters;
(2) Precision, namely the number of true heavy hitters reported over the number of answers reported;
(3) Average relative error of the reported frequencies, measured separately for the true heavy hitters and the false positive answers.

We perform the one-shot experiments using 
kosarak dataset\footnote{Kosarak. \url{http://fimi.ua.ac.be/data/}.}, which consists of 990002 click streams over 41270 different pages, and we uniformly partition the dataset across all parties. We use MurmurHash3 as the hash functions for count sketch. In our experiments, the default frequency threshold $\phi$ is $0.001$ (i.e. the items with frequency above $\phi N$ are defined as heavy hitters), the average message size $s$ is set to $3/(\phi \sqrt{k})$ (leading to an error guarantee of $\phi N / 3$), the default privacy parameter $\varepsilon$ is $2$, the default number of parties $k$ is $100$, the default number of rows of the count sketch is set to $5$, and all results are averaged over 5 repetitions.

\begin{figure}[htbp]
     \centering
     \begin{minipage}[t]{0.45\textwidth}
        \centering
        \includegraphics[width=\textwidth]{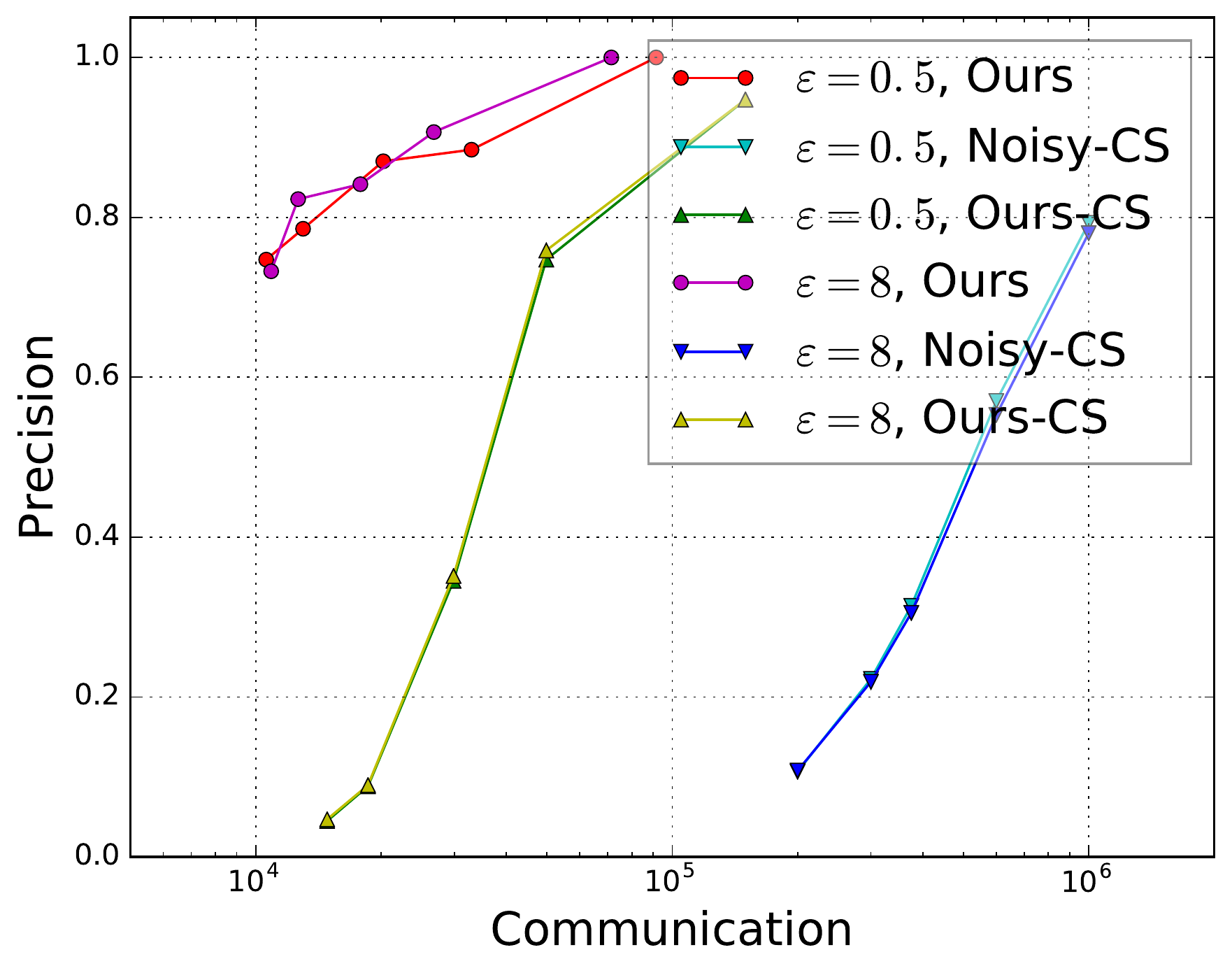}
        \subcaption{Precision}
        \label{fig:comm_precision}
     \end{minipage}
     \quad
     \begin{minipage}[t]{0.45\textwidth}
         \centering
         \includegraphics[width=\textwidth]{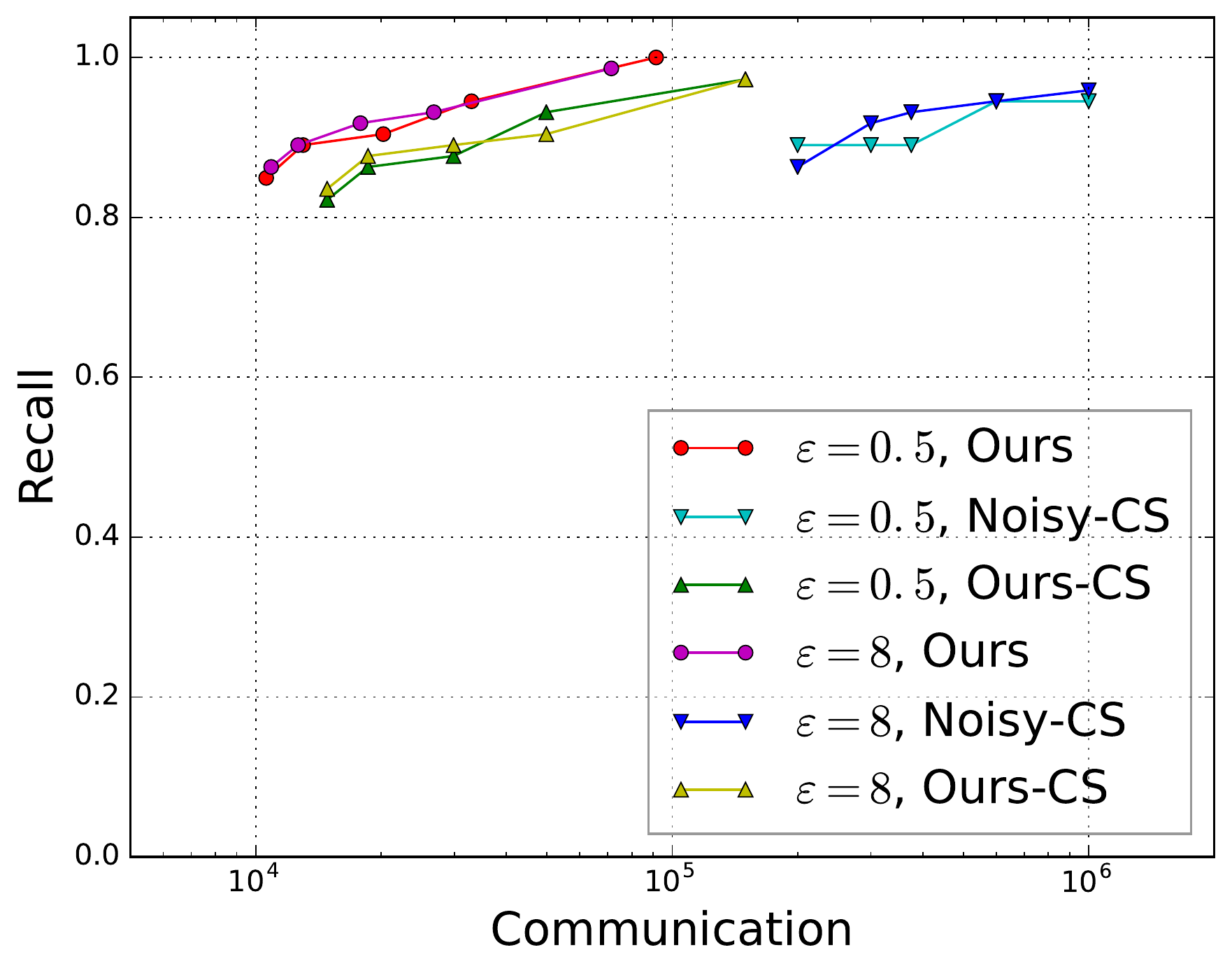}
         \subcaption{Recall}
         \label{fig:comm_recall}
     \end{minipage}
     \\
     \begin{minipage}[t]{0.45\textwidth}
        \centering
        \includegraphics[width=\textwidth]{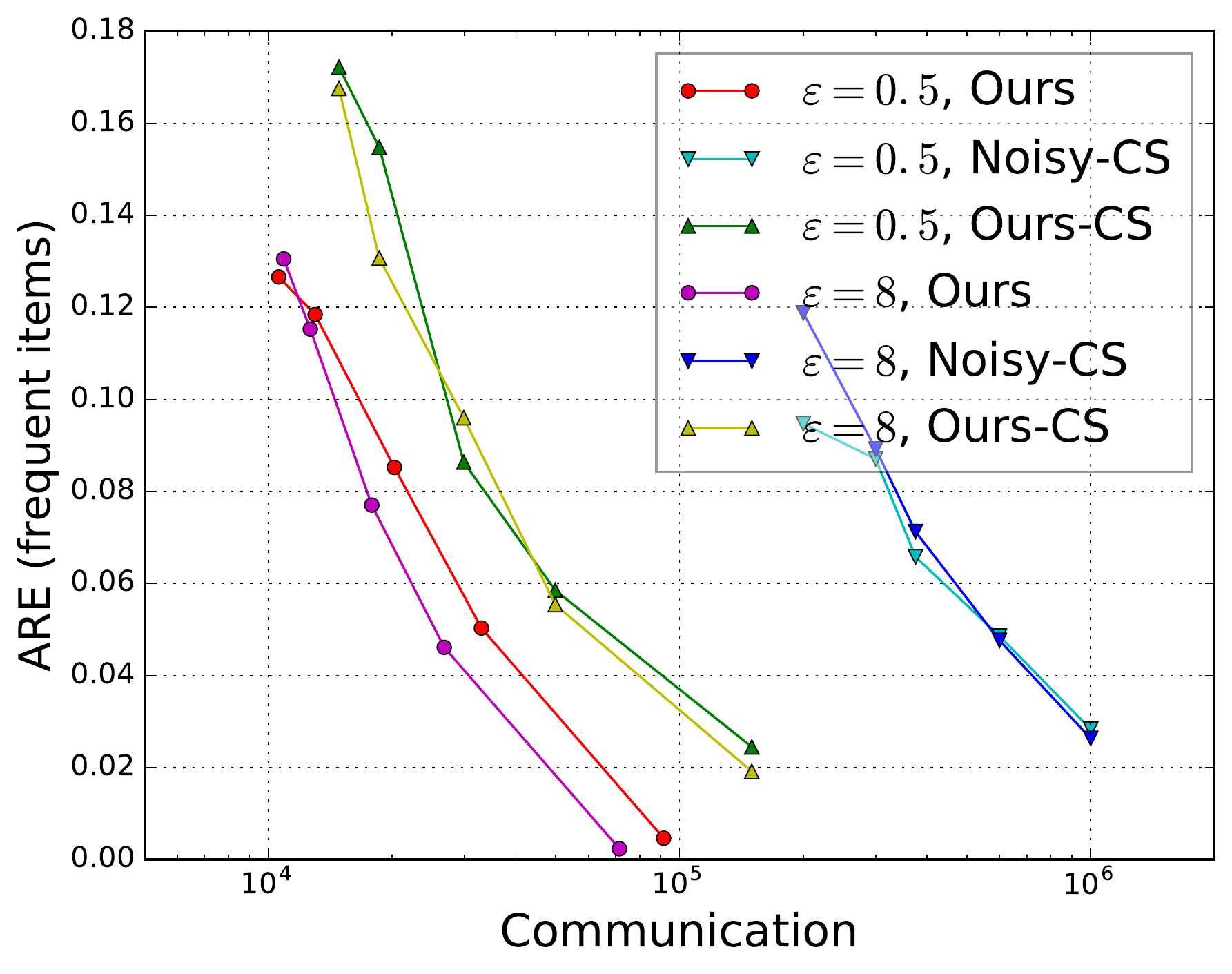}
        \subcaption{ARE (frequent items)}
        \label{fig:comm_are_fi}
     \end{minipage}
     \quad
     \begin{minipage}[t]{0.45\textwidth}
         \centering
         \includegraphics[width=\textwidth]{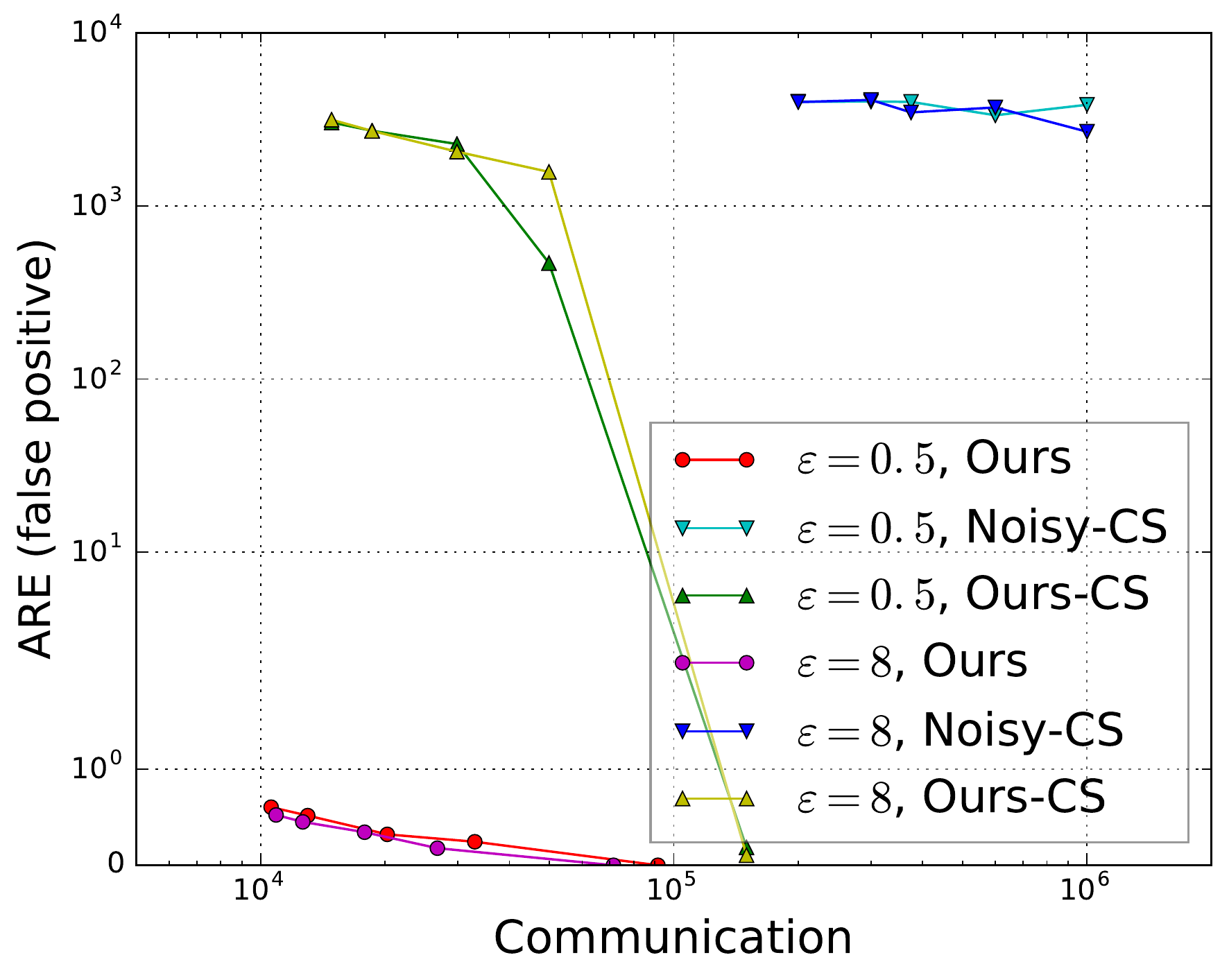}
         \subcaption{ARE (false positive)}
         \label{fig:comm_are_fp}
     \end{minipage}
     \caption{Accuracy vs. Communication.}
     \label{fig:acc_comm}
\end{figure}

We compare our methods with a baseline method (denoted as \texttt{Noisy-CS}): each party transmits a noisy count sketch of the same size $s$ to the aggregator, then the aggregator merges all noisy count sketches and takes the median estimator from all rows (following the convention of using linear sketches in the non-private setting). In Figure~\ref{fig:acc_comm}, we perform the experiments by varying $s$ and report the communication error trade-off, where \texttt{Ours-CS} denotes the method in Section~\ref{sec:one-shot-base} and \texttt{Ours} denotes the method in Section~\ref{sec:one-shot-hh}.

\begin{figure}[htbp]
  \centering
  \begin{minipage}[t]{0.45\textwidth}
  \centering \includegraphics[width=\textwidth]{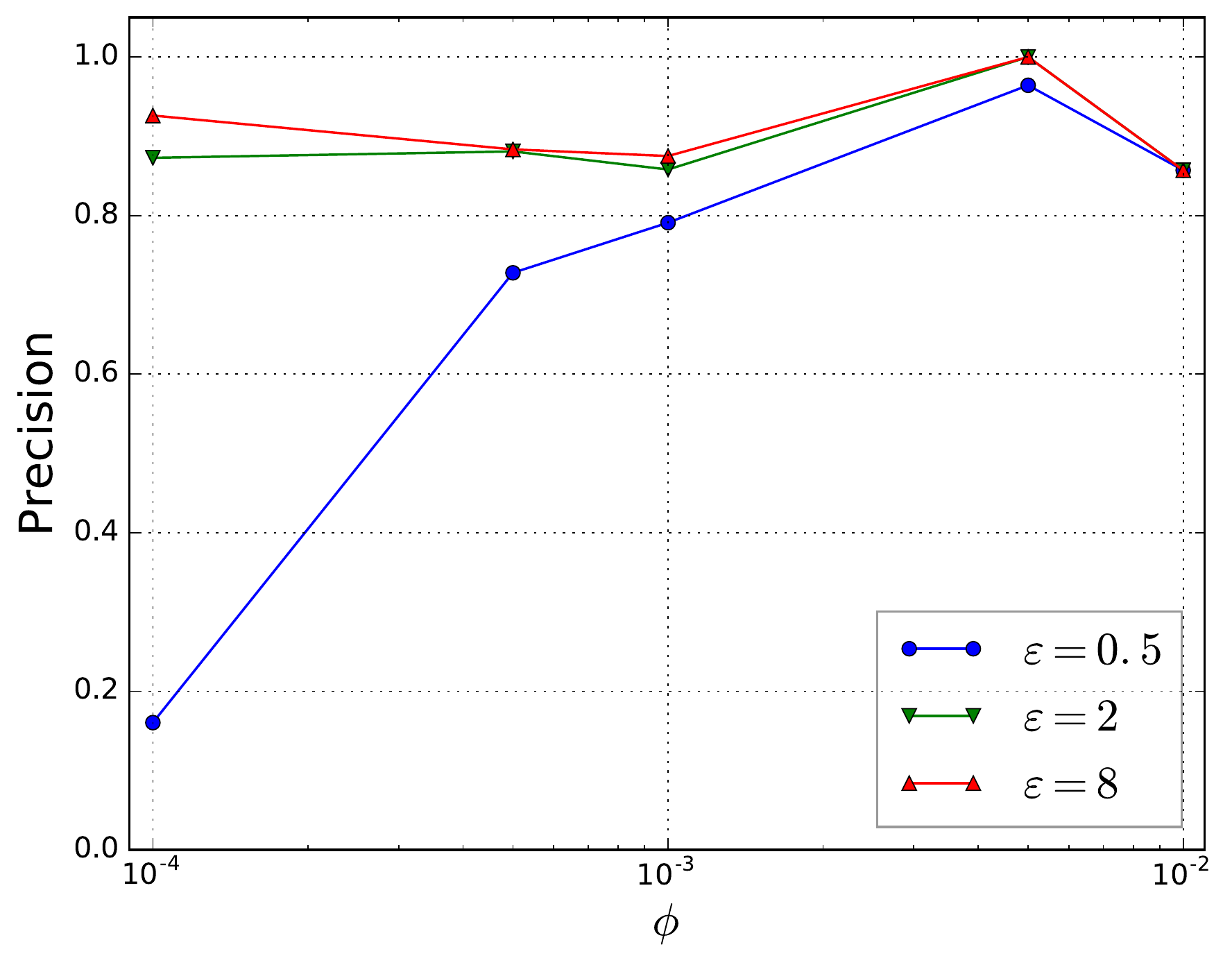}
  \subcaption{Precision}
  \label{fig:phi_precision}
  \end{minipage}
  \quad
  \begin{minipage}[t]{0.45\textwidth}
   \centering
   \includegraphics[width=\textwidth]{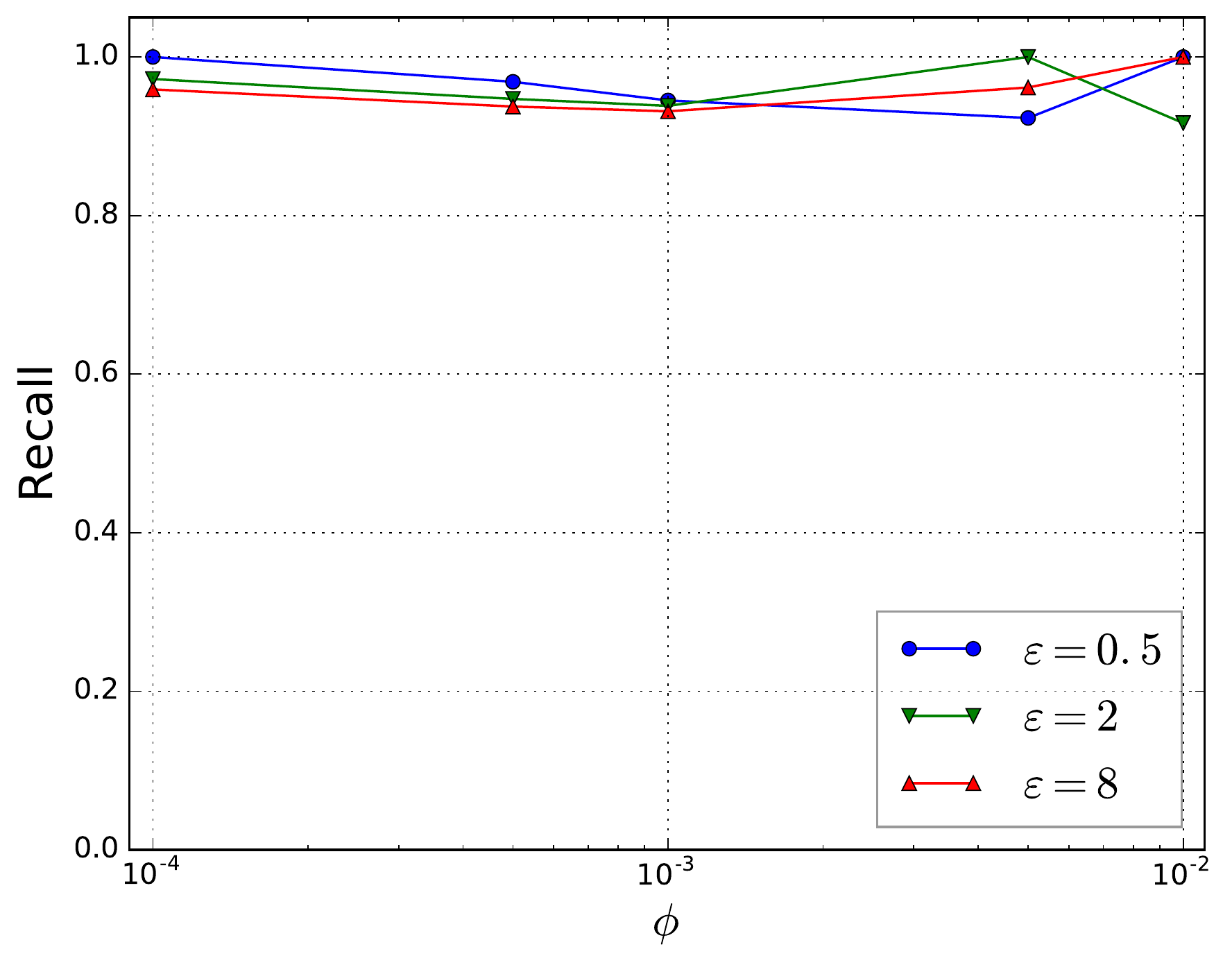}
   \subcaption{Recall}
   \label{fig:phi_recall}
  \end{minipage}
  \\
  \begin{minipage}[t]{0.45\textwidth}
  \centering \includegraphics[width=\textwidth]{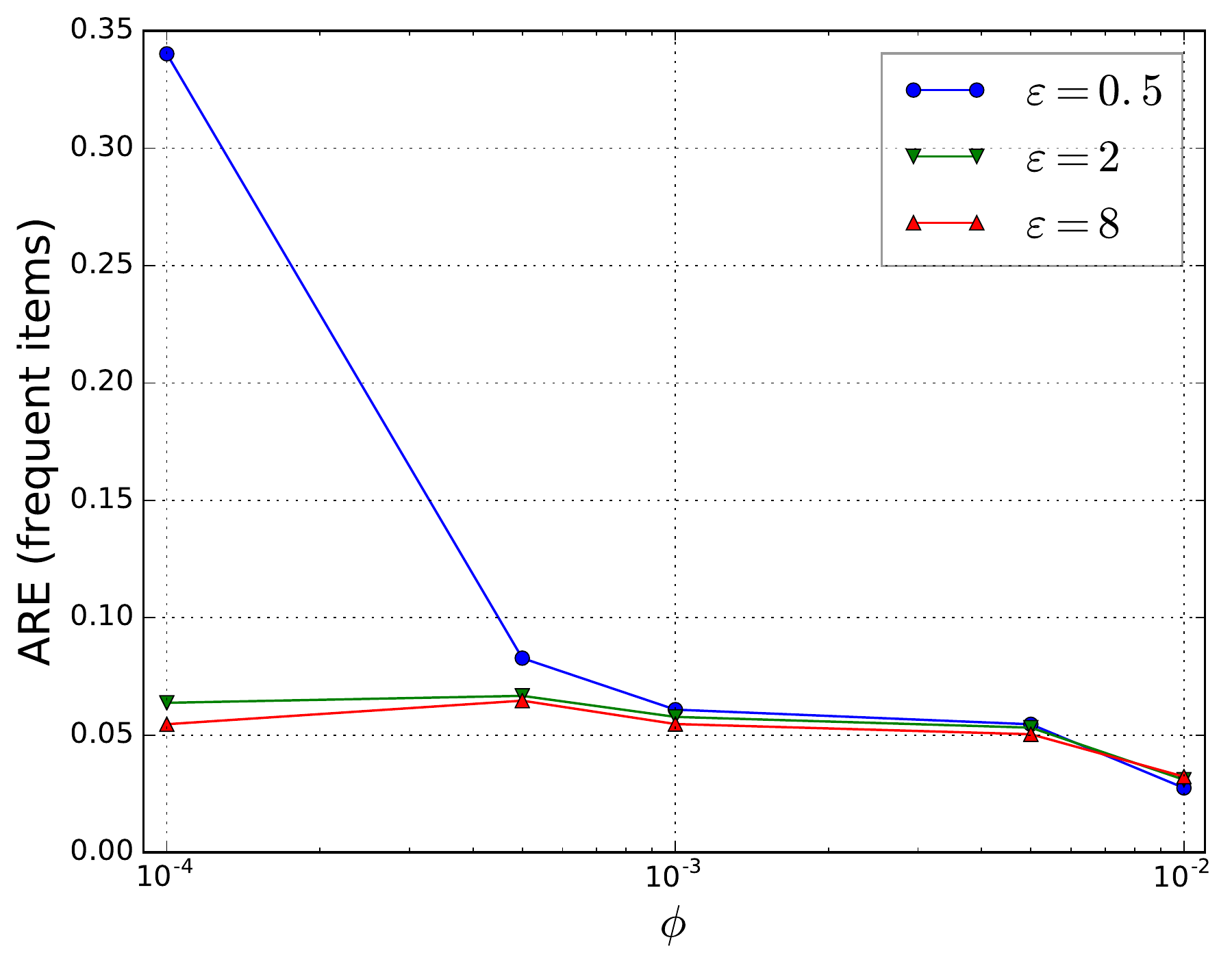}
  \subcaption{ARE (frequent items)}
  \label{fig:phi_are_fi}
  \end{minipage}
  \quad
  \begin{minipage}[t]{0.45\textwidth}
   \centering
   \includegraphics[width=\textwidth]{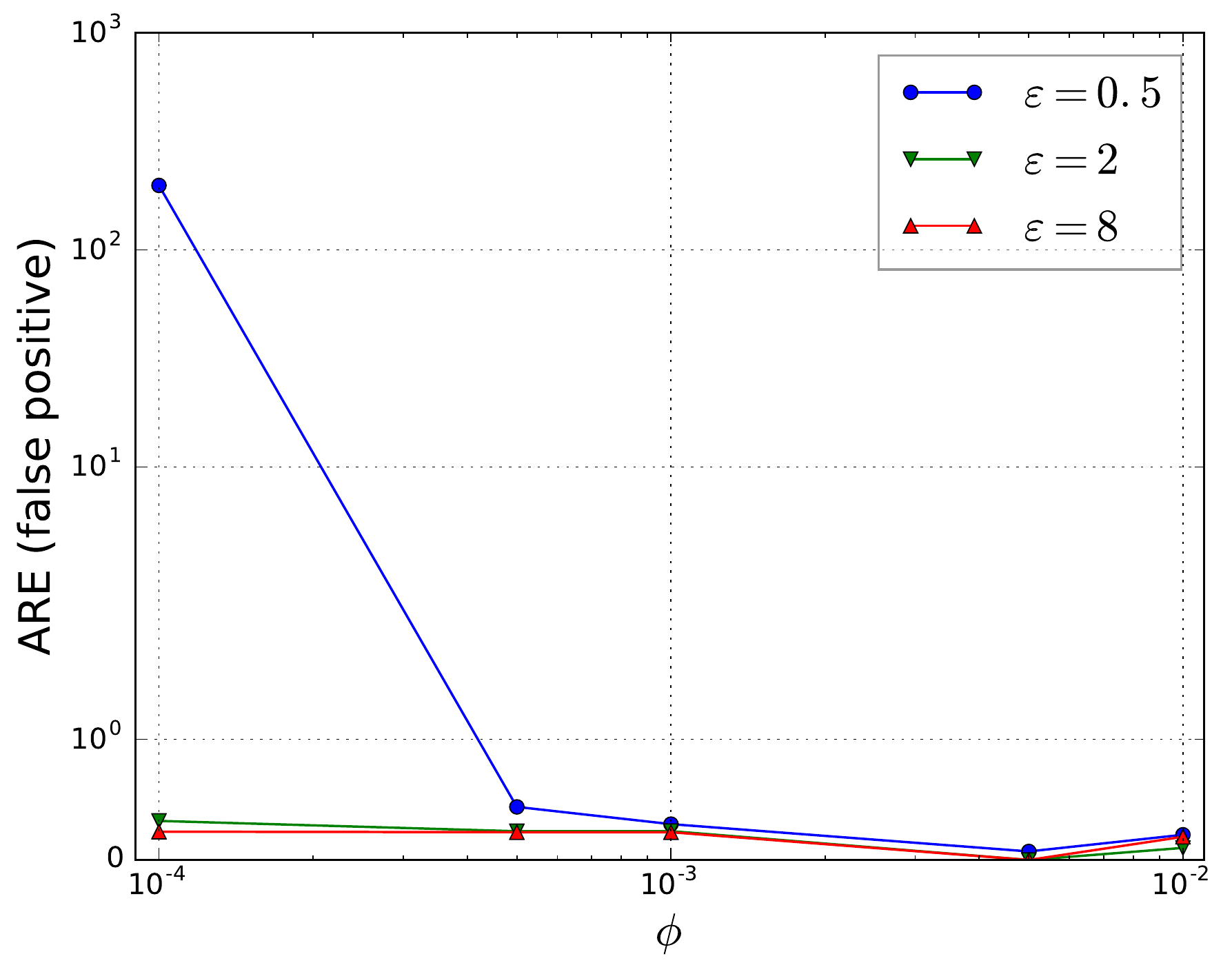}
   \subcaption{ARE (false positive)}
   \label{fig:phi_are_fp}
  \end{minipage}
  \caption{Accuracy vs. $\phi$.}
  \label{fig:acc_phi}
\end{figure}

Figure~\ref{fig:acc_phi} shows results for varying $\phi$. It is shown that our method has high accuracy in practice for various frequency thresholds.

\begin{figure}[htbp]
     \centering
     \begin{minipage}[t]{0.42\textwidth}
        \centering
        \includegraphics[width=\textwidth]{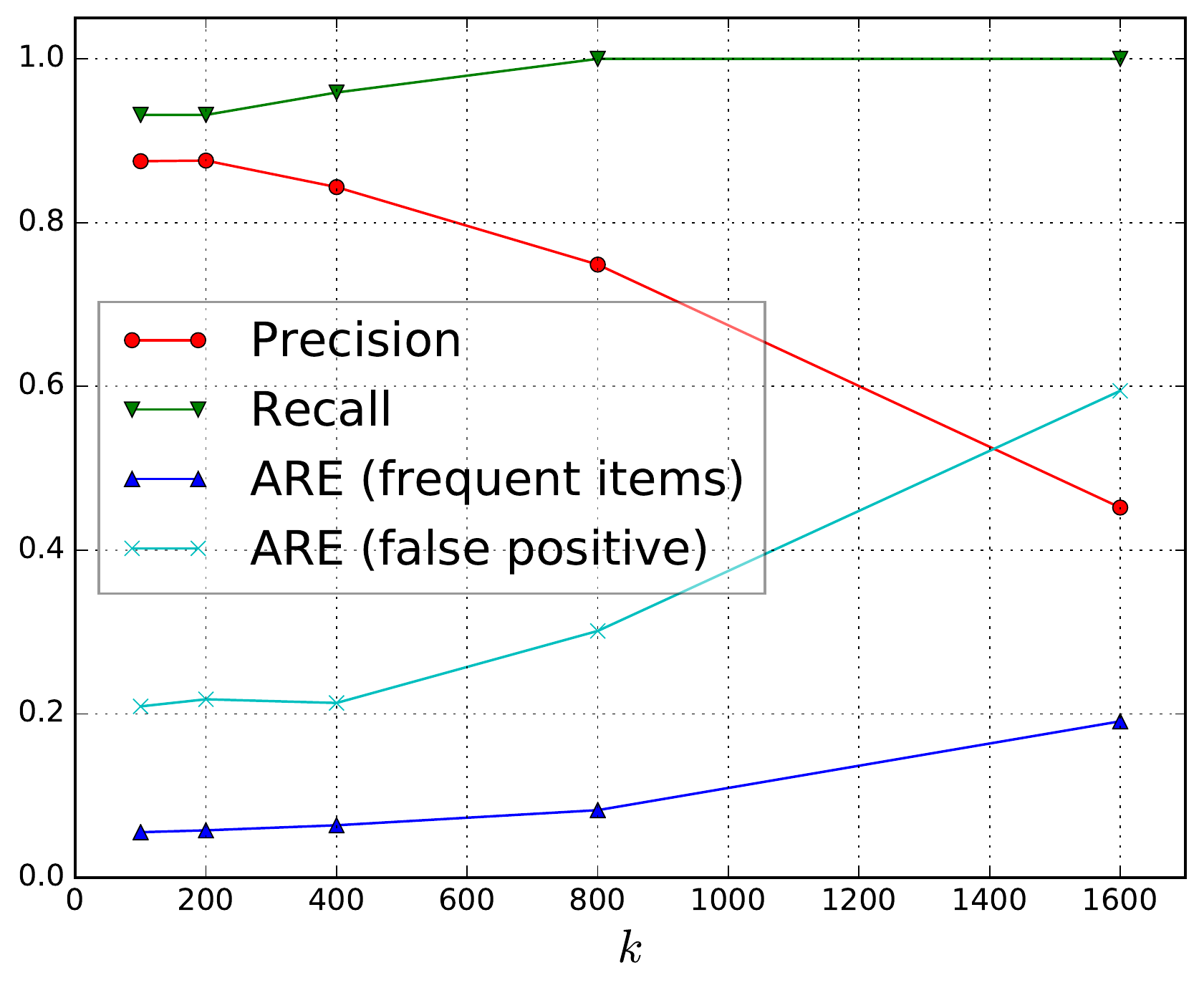}
        \subcaption{Accuracy vs. $k$}
        \label{fig:acc_k}
     \end{minipage}
     \quad
     \begin{minipage}[t]{0.45\textwidth}
         \centering
         \includegraphics[width=\textwidth]{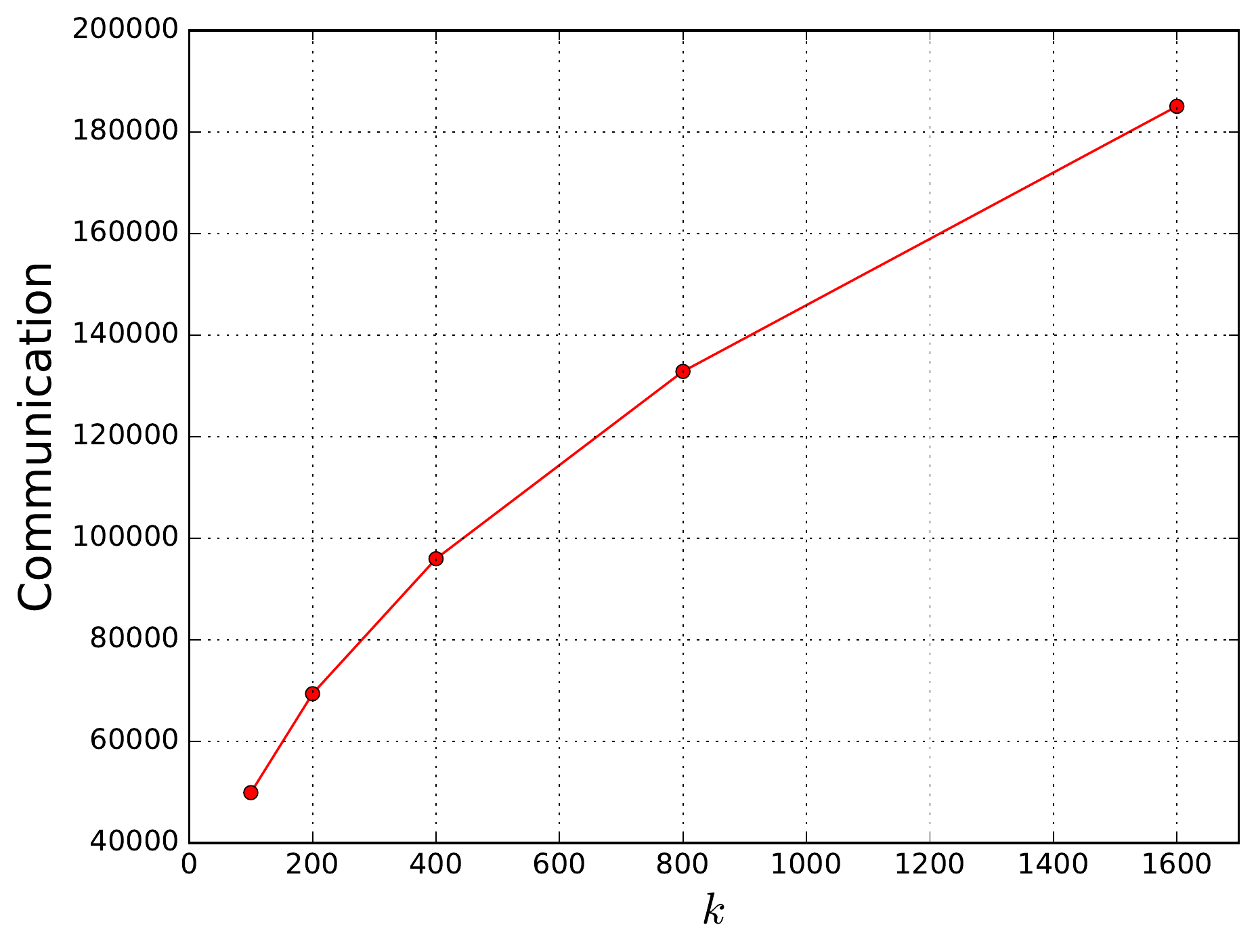}
         \subcaption{Communication vs. $k$}
         \label{fig:comm_k}
     \end{minipage}

     \caption{Accuracy/Communication vs. $k$.}
     \label{fig:metric_k}
\end{figure}

Figure~\ref{fig:metric_k} is performed by varying $k$. We also plot the curve of communication vs. $k$. The accuracy slightly degrades as $k$ increases, since the error ($\sqrt{k} / \varepsilon$) due to privacy constraint becomes larger.

\subsection{Streaming}

\begin{figure}[htbp]
     \centering
     \begin{minipage}[t]{0.45\textwidth}
        \centering
        \includegraphics[width=\textwidth]{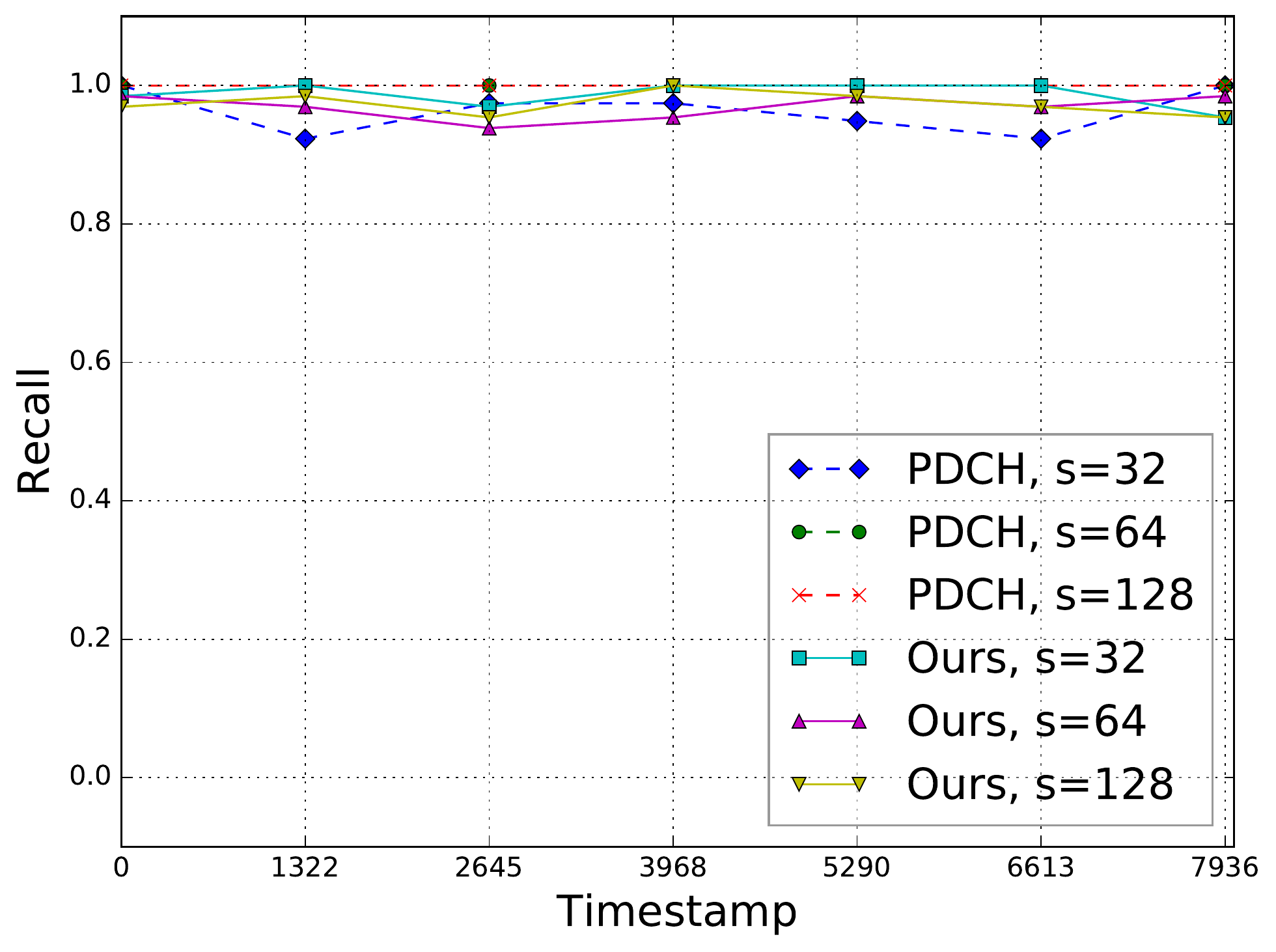}
        \subcaption{Recall}
        \label{fig:recall}
     \end{minipage}
     \quad
     \begin{minipage}[t]{0.45\textwidth}
         \centering
         \includegraphics[width=\textwidth]{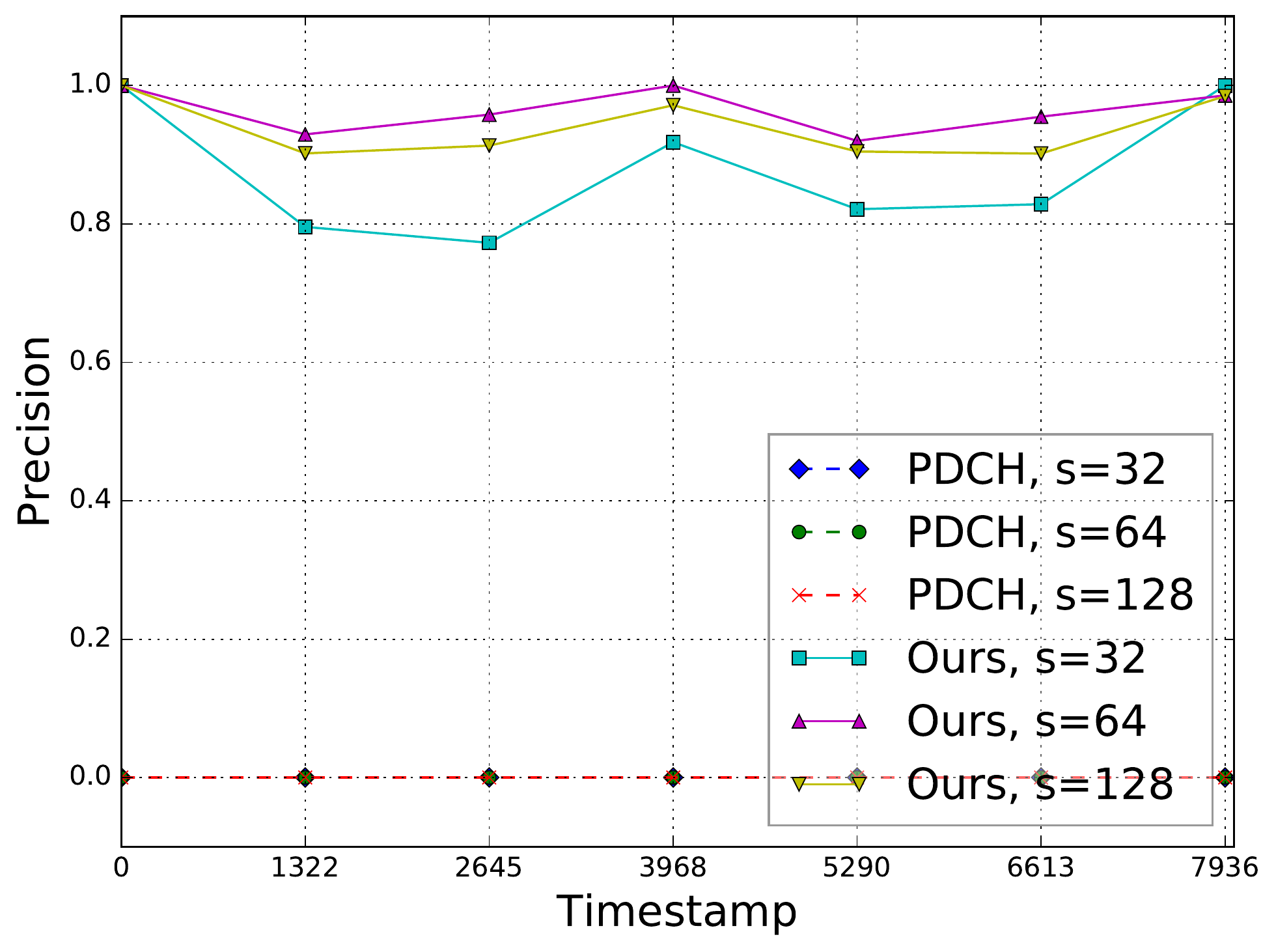}
         \subcaption{Precision}
         \label{fig:precision}
     \end{minipage}
     \\
     \begin{minipage}[t]{0.45\textwidth}
         \centering
         \includegraphics[width=\textwidth]{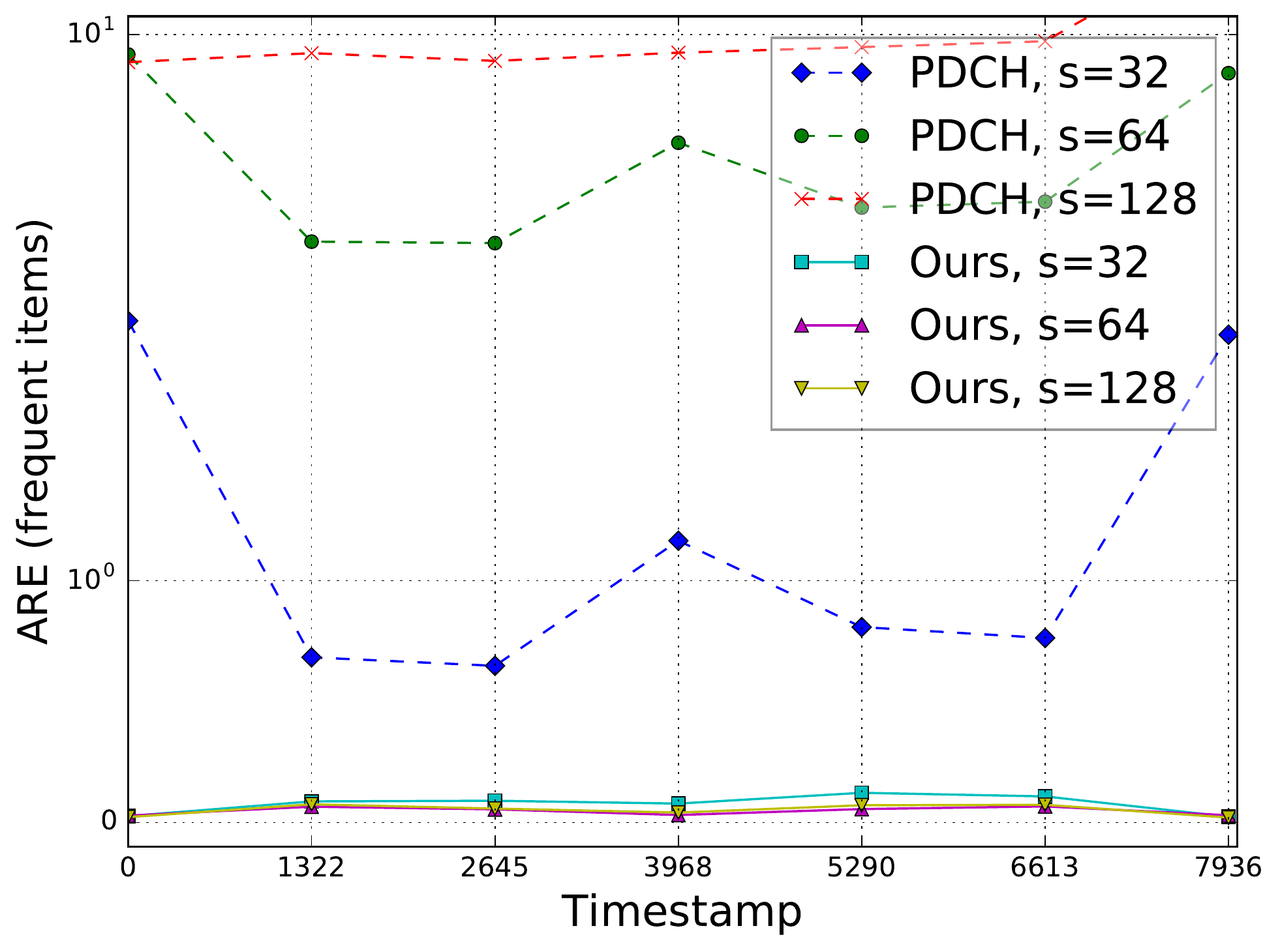}
         \subcaption{ARE (frequent items)}
         \label{fig:are_t}
     \end{minipage}
     \quad
      \begin{minipage}[t]{0.45\textwidth}
         \centering
         \includegraphics[width=\textwidth]{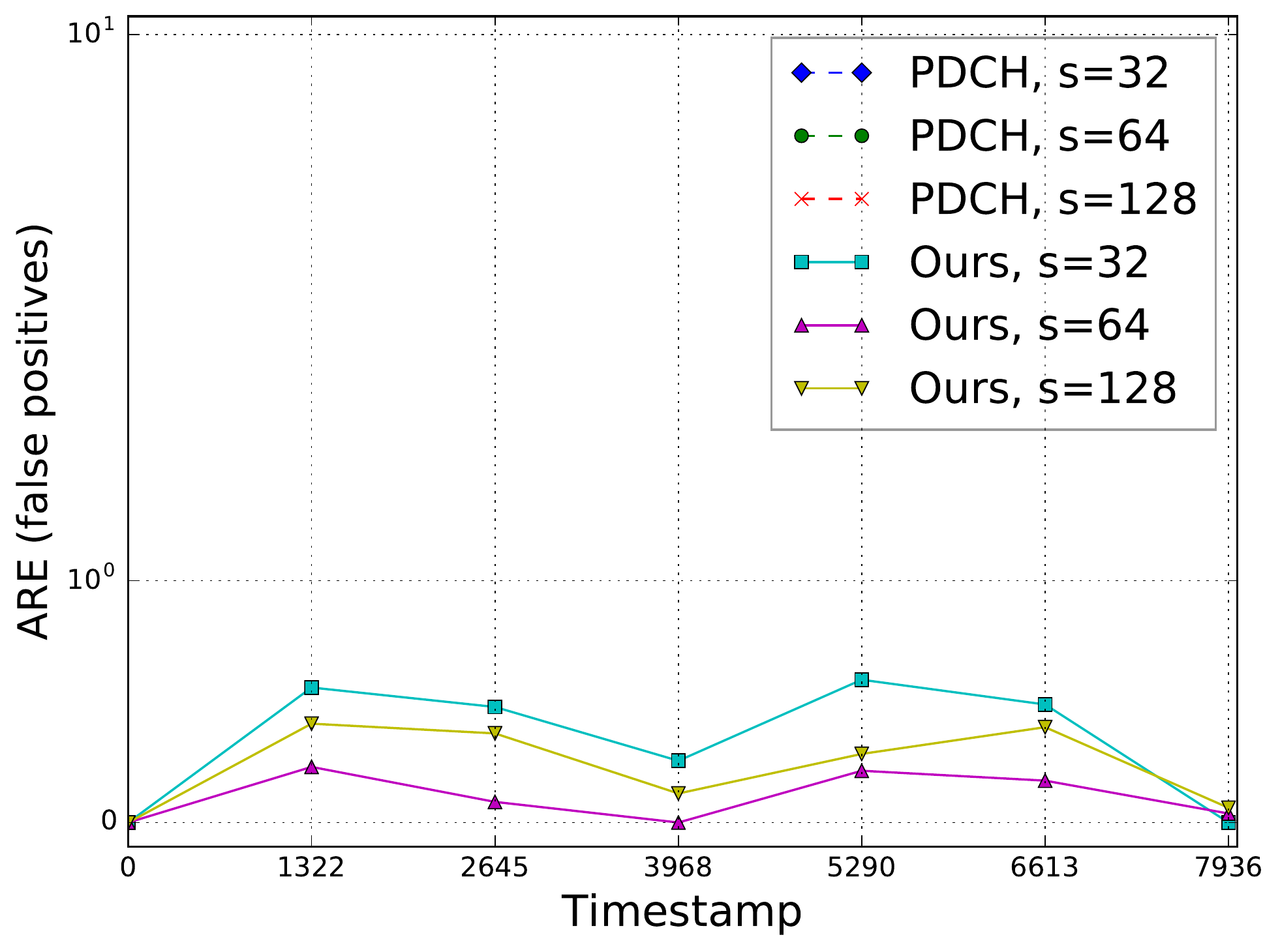}
         \subcaption{ARE (false positive)}
         \label{fig:are_f}
     \end{minipage}
     \caption{Accuracy over time.}
     \label{fig:conti_ours}
\end{figure}

We evaluate our sliding window protocol on the kosarak dataset.
By default we set the number of parties $k=100$, the privacy parameter $\varepsilon=4$, the frequency threshold $\phi=0.005$, the number of rows of count sketch is 5, and the results are averaged over 5 independent repetitions.
The window size is set to $w=n/10$, where $n$ is the length of the stream.

\begin{figure}[htbp]
     \centering
       \begin{minipage}[t]{0.45\textwidth}
         \centering
         \includegraphics[width=\textwidth]{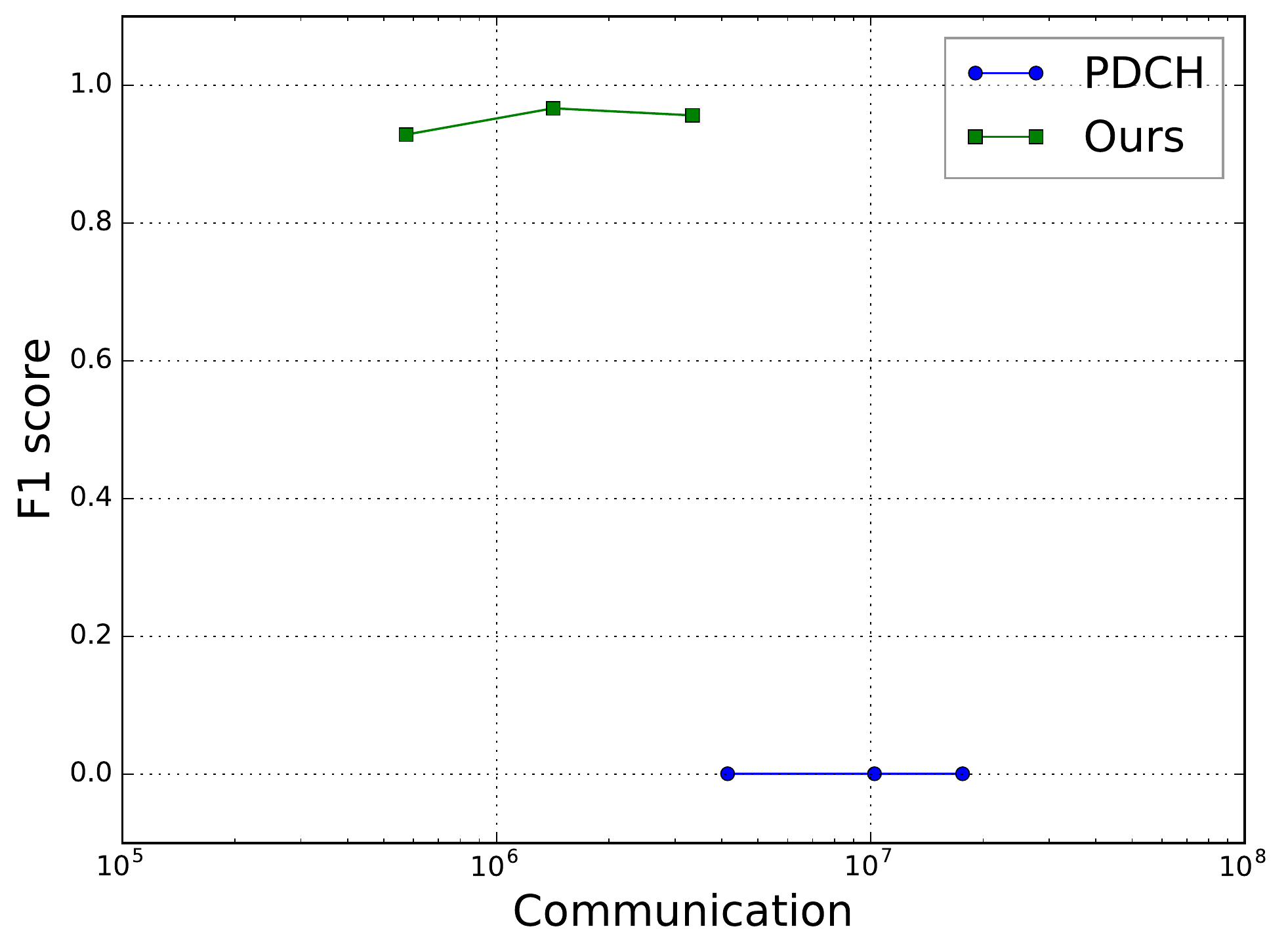}
         \subcaption{F1 vs Communication}
         \label{fig:commu_f1}
     \end{minipage}
     \quad
     \begin{minipage}[t]{0.45\textwidth}
         \centering
         \includegraphics[width=\textwidth]{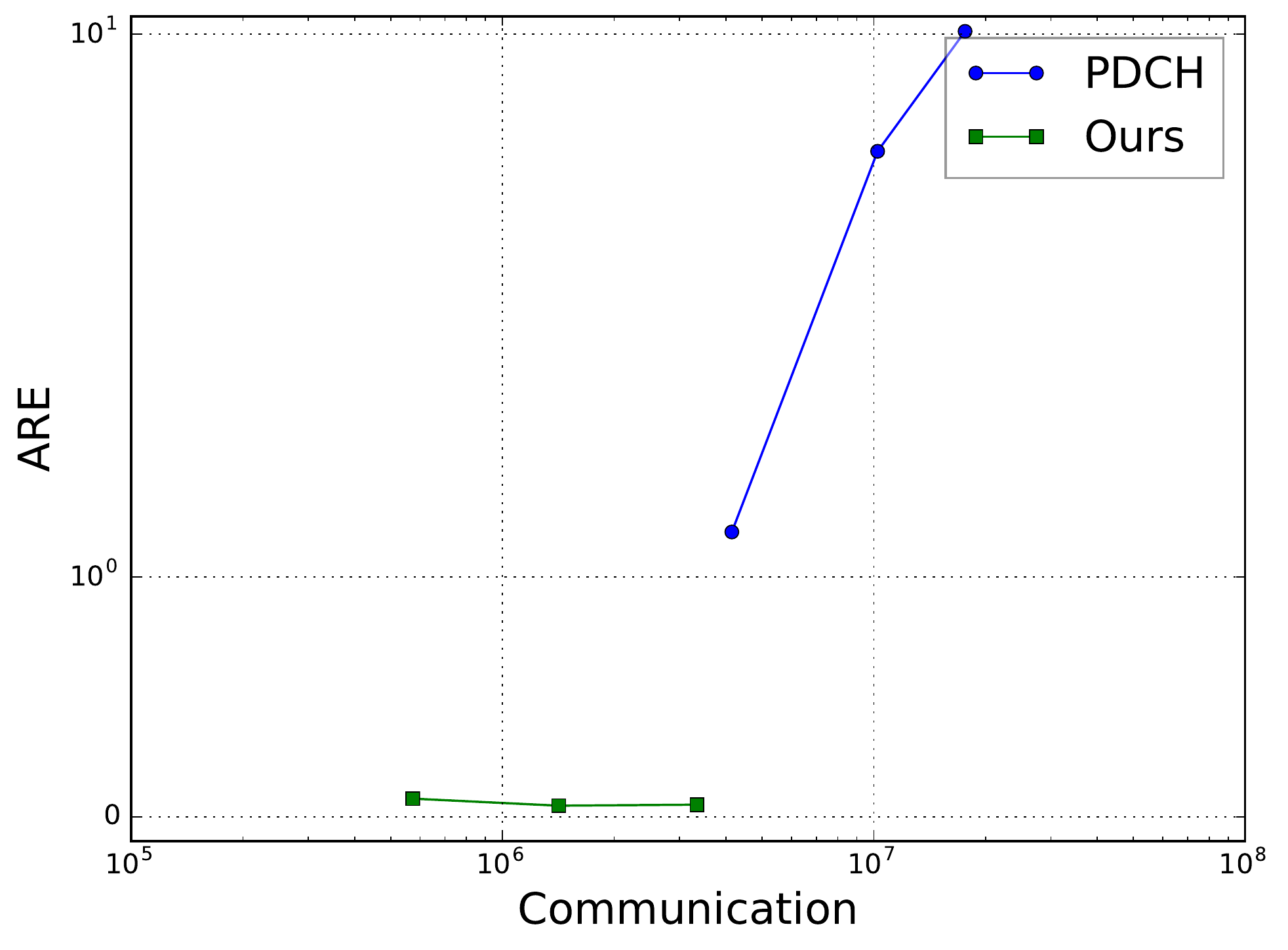}
         \subcaption{ARE vs Communication}
         \label{fig:commu_are}
     \end{minipage}
     \caption{Accuracy vs. communication}
     \label{fig:conti_ours_comm}
\end{figure}

We compare our method with the PDCH protocol \cite{chan2012differentially} using the same $s$, which leads to the same asymptotic communication bound.
We report the accuracy (precision/recall/ARE) over time in Figure~\ref{fig:conti_ours}, and the accuracy (F1 score and ARE) vs. communication cost in Figure~\ref{fig:conti_ours_comm}.
Observe that increasing $s$ does not always result in better accuracy in practice.
In our method, this is because increasing $s$ also leads to more epochs, making the noise magnitude required in the inter-epoch part larger. In PDCH, the accuracy may drop significantly when increasing $s$, due to the large sensitivity of the PMG summary used in their protocol. More precisely, for a MG summary of size $s$, PMG injects noises of magnitude $\tilde{\Theta}(s/\varepsilon)$ onto it to preserve differential privacy. In practice, the noises can even be orders of magnitudes larger than the real frequencies, making PDCH impractical. Specifically, in our setting, the noise in PDCH is roughly 10 times larger than the real frequencies, so a 0-frequency item can be easily reported as frequent item, making the ARE (false positive) infinite.

}{We evaluated our one-shot and streaming protocols on the Kosarak dataset\footnote{Kosarak. \url{http://fimi.ua.ac.be/data/}}, and the results are presented in the supplementary material.}

\bibliography{paper}

\begin{thebibliography}{10}

\bibitem{acharya2019communication}
Jayadev Acharya and Ziteng Sun.
\newblock Communication complexity in locally private distribution estimation
  and heavy hitters.
\newblock In {\em International Conference on Machine Learning}, pages 51--60.
  PMLR, 2019.

\bibitem{acharya2019hadamard}
Jayadev Acharya, Ziteng Sun, and Huanyu Zhang.
\newblock Hadamard response: Estimating distributions privately, efficiently,
  and with little communication.
\newblock In {\em The 22nd International Conference on Artificial Intelligence
  and Statistics}, pages 1120--1129, 2019.

\bibitem{agarwal2018cpsgd}
Naman Agarwal, Ananda~Theertha Suresh, Felix Xinnan~X Yu, Sanjiv Kumar, and
  Brendan McMahan.
\newblock cpsgd: Communication-efficient and differentially-private distributed
  sgd.
\newblock In {\em Advances in Neural Information Processing Systems}, pages
  7564--7575, 2018.

\bibitem{appledp}
Apple.
\newblock Apple differential privacy technical overview.
\newblock https://www.apple.com/privacy/docs/Differential Privacy Overview.pdf,
  2017.

\bibitem{balle2019privacy}
Borja Balle, James Bell, Adri{\`a} Gasc{\'o}n, and Kobbi Nissim.
\newblock The privacy blanket of the shuffle model.
\newblock In {\em Annual International Cryptology Conference}, pages 638--667.
  Springer, 2019.

\bibitem{bassily2017practical}
Raef Bassily, Kobbi Nissim, Uri Stemmer, and Abhradeep Thakurta.
\newblock Practical locally private heavy hitters.
\newblock In {\em Proceedings of the 31st International Conference on Neural
  Information Processing Systems}, pages 2285--2293, 2017.

\bibitem{bassily2020practical}
Raef Bassily, Kobbi Nissim, Uri Stemmer, and Abhradeep Thakurta.
\newblock Practical locally private heavy hitters.
\newblock {\em J. Mach. Learn. Res.}, 21:16:1--16:42, 2020.

\bibitem{bassily2015local}
Raef Bassily and Adam Smith.
\newblock Local, private, efficient protocols for succinct histograms.
\newblock In {\em Proceedings of the forty-seventh annual ACM symposium on
  Theory of computing}, pages 127--135, 2015.

\bibitem{beimel2008distributed}
Amos Beimel, Kobbi Nissim, and Eran Omri.
\newblock Distributed private data analysis: Simultaneously solving how and
  what.
\newblock In {\em Annual International Cryptology Conference}, pages 451--468.
  Springer, 2008.

\bibitem{chakrabarti2015data}
Amit Chakrabarti.
\newblock Data stream algorithms.
\newblock {\em Computer Science}, 49:149, 2015.

\bibitem{chan2012differentially}
T-H~Hubert Chan, Mingfei Li, Elaine Shi, and Wenchang Xu.
\newblock Differentially private continual monitoring of heavy hitters from
  distributed streams.
\newblock In {\em International Symposium on Privacy Enhancing Technologies
  Symposium}, pages 140--159. Springer, 2012.

\bibitem{chan2011private}
T-H~Hubert Chan, Elaine Shi, and Dawn Song.
\newblock Private and continual release of statistics.
\newblock {\em ACM Transactions on Information and System Security (TISSEC)},
  14(3):1--24, 2011.

\bibitem{chan2012optimal}
TH~Hubert Chan, Elaine Shi, and Dawn Song.
\newblock Optimal lower bound for differentially private multi-party
  aggregation.
\newblock In {\em European Symposium on Algorithms}, pages 277--288. Springer,
  2012.

\bibitem{charikar2002finding}
Moses Charikar, Kevin Chen, and Martin Farach-Colton.
\newblock Finding frequent items in data streams.
\newblock In {\em International Colloquium on Automata, Languages, and
  Programming}, pages 693--703. Springer, 2002.

\bibitem{chen2020breaking}
Wei-Ning Chen, Peter Kairouz, and Ayfer {\"O}zg{\"u}r.
\newblock Breaking the communication-privacy-accuracy trilemma.
\newblock In {\em Advances in Neural Information Processing Systems}, 2020.

\bibitem{cormode2010methods}
Graham Cormode and Marios Hadjieleftheriou.
\newblock Methods for finding frequent items in data streams.
\newblock {\em The VLDB Journal}, 19(1):3--20, 2010.

\bibitem{cormode2018tutorial}
Graham Cormode, Somesh Jha, Tejas Kulkarni, Ninghui Li, Divesh Srivastava, and
  Tianhao Wang.
\newblock Privacy at scale: Local differential privacy in practice.
\newblock In {\em Proceedings of the 2018 International Conference on
  Management of Data}, pages 1655--1658. {ACM}, 2018.

\bibitem{cormode2019answering}
Graham Cormode, Tejas Kulkarni, and Divesh Srivastava.
\newblock Answering range queries under local differential privacy.
\newblock {\em Proceedings of the VLDB Endowment}, 12(10):1126--1138, 2019.

\bibitem{cormode11functional}
Graham Cormode, Senthilmurugan Muthukrishnan, and Ke~Yi.
\newblock Algorithms for distributed functional monitoring.
\newblock {\em ACM Transactions on Algorithms}, 7:21, 03 2011.

\bibitem{cormode12continuous}
Graham Cormode, Senthilmurugan Muthukrishnan, Ke~Yi, and Qin Zhang.
\newblock Continuous sampling from distributed streams.
\newblock {\em Journal of The ACM}, 59:1--25, 04 2012.

\bibitem{cormode2005improved}
Graham Cormode and Shan Muthukrishnan.
\newblock An improved data stream summary: the count-min sketch and its
  applications.
\newblock {\em Journal of Algorithms}, 55(1):58--75, 2005.

\bibitem{cormode2020book}
Graham Cormode and Ke~Yi.
\newblock {\em Small Summaries for Big Data}.
\newblock Cambridge University Press, 2020.

\bibitem{ding2017collecting}
Bolin Ding, Janardhan Kulkarni, and Sergey Yekhanin.
\newblock Collecting telemetry data privately.
\newblock In {\em Advances in Neural Information Processing Systems}, pages
  3571--3580, 2017.

\bibitem{dwork2006our}
Cynthia Dwork, Krishnaram Kenthapadi, Frank McSherry, Ilya Mironov, and Moni
  Naor.
\newblock Our data, ourselves: Privacy via distributed noise generation.
\newblock In {\em Annual International Conference on the Theory and
  Applications of Cryptographic Techniques}, pages 486--503. Springer, 2006.

\bibitem{dwork2010differential}
Cynthia Dwork, Moni Naor, Toniann Pitassi, and Guy~N Rothblum.
\newblock Differential privacy under continual observation.
\newblock In {\em Proceedings of the forty-second ACM symposium on Theory of
  computing}, pages 715--724, 2010.

\bibitem{dwork2014algorithmic}
Cynthia Dwork, Aaron Roth, et~al.
\newblock The algorithmic foundations of differential privacy.
\newblock {\em Foundations and Trends{\textregistered} in Theoretical Computer
  Science}, 9(3--4):211--407, 2014.

\bibitem{erlingsson2019shuffling}
{\'{U}}lfar Erlingsson, Vitaly Feldman, Ilya Mironov, Ananth Raghunathan, Kunal
  Talwar, and Abhradeep Thakurta.
\newblock Amplification by shuffling: From local to central differential
  privacy via anonymity.
\newblock In {\em Proceedings of the Thirtieth Annual {ACM-SIAM} Symposium on
  Discrete Algorithms}, pages 2468--2479. {SIAM}, 2019.

\bibitem{erlingsson2014rappor}
{\'{U}}lfar Erlingsson, Vasyl Pihur, and Aleksandra Korolova.
\newblock {RAPPOR:} randomized aggregatable privacy-preserving ordinal
  response.
\newblock In Gail{-}Joon Ahn, Moti Yung, and Ninghui Li, editors, {\em
  Proceedings of the 2014 {ACM} {SIGSAC} Conference on Computer and
  Communications Security}, pages 1054--1067. {ACM}, 2014.

\bibitem{feldman2021lossless}
Vitaly Feldman and Kunal Talwar.
\newblock Lossless compression of efficient private local randomizers.
\newblock {\em arXiv preprint arXiv:2102.12099}, 2021.

\bibitem{ghazi2019power}
Badih Ghazi, Noah Golowich, Ravi Kumar, Rasmus Pagh, and Ameya Velingker.
\newblock On the power of multiple anonymous messages.
\newblock {\em arXiv preprint arXiv:1908.11358}, 2019.

\bibitem{ghazi2020private}
Badih Ghazi, Ravi Kumar, Pasin Manurangsi, and Rasmus Pagh.
\newblock Private counting from anonymous messages: Near-optimal accuracy with
  vanishing communication overhead.
\newblock In {\em International Conference on Machine Learning}, pages
  3505--3514. PMLR, 2020.

\bibitem{ghosh2012universally}
Arpita Ghosh, Tim Roughgarden, and Mukund Sundararajan.
\newblock Universally utility-maximizing privacy mechanisms.
\newblock {\em SIAM Journal on Computing}, 41(6):1673--1693, 2012.

\bibitem{hamm2016learning}
Jihun Hamm, Yingjun Cao, and Mikhail Belkin.
\newblock Learning privately from multiparty data.
\newblock In {\em International Conference on Machine Learning}, pages
  555--563. PMLR, 2016.

\bibitem{huang2017communication}
Zengfeng Huang and Ke~Yi.
\newblock The communication complexity of distributed epsilon-approximations.
\newblock {\em SIAM Journal on Computing}, 46(4):1370--1394, 2017.

\bibitem{kairouz2019advances}
Peter Kairouz, H~Brendan McMahan, Brendan Avent, Aur{\'e}lien Bellet, Mehdi
  Bennis, Arjun~Nitin Bhagoji, Keith Bonawitz, Zachary Charles, Graham Cormode,
  Rachel Cummings, et~al.
\newblock Advances and open problems in federated learning.
\newblock {\em arXiv preprint arXiv:1912.04977}, 2019.

\bibitem{kairouz2014extremal}
Peter Kairouz, Sewoong Oh, and Pramod Viswanath.
\newblock Extremal mechanisms for local differential privacy.
\newblock {\em Advances in neural information processing systems},
  27:2879--2887, 2014.

\bibitem{matouvsek2001probabilistic}
Ji{\v{r}}{\'\i} Matou{\v{s}}ek and Jan Vondr{\'a}k.
\newblock The probabilistic method.
\newblock {\em Lecture Notes, Department of Applied Mathematics, Charles
  University, Prague}, 2001.

\bibitem{pathak2010multiparty}
Manas Pathak, Shantanu Rane, and Bhiksha Raj.
\newblock Multiparty differential privacy via aggregation of locally trained
  classifiers.
\newblock In {\em Advances in Neural Information Processing Systems}, pages
  1876--1884, 2010.

\bibitem{shi17coalition}
Elaine Shi, T.-H Chan, Eleanor Rieffel, and Dawn Song.
\newblock Distributed private data analysis: Lower bounds and practical
  constructions.
\newblock {\em ACM Transactions on Algorithms}, 13:1--38, 12 2017.

\bibitem{upadhyay2019sublinear}
Jalaj Upadhyay.
\newblock Sublinear space private algorithms under the sliding window model.
\newblock In {\em International Conference on Machine Learning}, pages
  6363--6372. PMLR, 2019.

\bibitem{vadhan2017complexity}
Salil Vadhan.
\newblock The complexity of differential privacy.
\newblock In {\em Tutorials on the Foundations of Cryptography}, pages
  347--450. Springer, 2017.

\bibitem{wang2013quantiles}
Lu~Wang, Ge~Luo, Ke~Yi, and Graham Cormode.
\newblock Quantiles over data streams: an experimental study.
\newblock In {\em Proceedings of the 2013 ACM SIGMOD International Conference
  on Management of Data}, pages 737--748, 2013.

\bibitem{wang2020survey}
Teng Wang, Xuefeng Zhang, Jingyu Feng, and Xinyu Yang.
\newblock A comprehensive survey on local differential privacy toward data
  statistics and analysis in crowdsensing.
\newblock {\em CoRR}, abs/2010.05253, 2020.

\bibitem{wang2017locally}
Tianhao Wang, Jeremiah Blocki, Ninghui Li, and Somesh Jha.
\newblock Locally differentially private protocols for frequency estimation.
\newblock In {\em 26th $\{$USENIX$\}$ Security Symposium ($\{$USENIX$\}$
  Security 17)}, pages 729--745, 2017.

\bibitem{yang2020survey}
Mengmeng Yang, Lingjuan Lyu, Jun Zhao, Tianqing Zhu, and Kwok{-}Yan Lam.
\newblock Local differential privacy and its applications: {A} comprehensive
  survey.
\newblock {\em CoRR}, abs/2008.03686, 2020.

\end{thebibliography}
\bibliographystyle{plain}

\end{document}